\documentstyle[11pt,epsfig]{article}
\newcommand \be{\begin{equation}}
\newcommand \ba{\begin{eqnarray}}
\newcommand \ee{\end{equation}}
\newcommand \ea{\end{eqnarray}}
\newcommand{\lp}{\left(}
\newcommand{\rp}{\right)}
\newcommand{\dj}{Dow Jones Industrial Average}  

\newcommand{\dds}{drawdowns }

[1999/03/29 PSNFSS v.7.2
Times font as default roman
: S Rahtz]

\begin{document}

\title{Large Stock Market Price Drawdowns Are Outliers}
\thispagestyle{empty}

\author{Anders Johansen$^1$ and Didier Sornette$^{1,2,3}$\\
$^1$ Institute of Geophysics and
Planetary Physics\\ University of California, Los Angeles, California 90095\\
$^2$ Department of Earth and Space Science\\
University of California, Los Angeles, California 90095\\
$^3$ Laboratoire de Physique de la Mati\`{e}re Condens\'{e}e\\ CNRS UMR6622 and
Universit\'{e} de Nice-Sophia Antipolis\\ B.P. 71, Parc
Valrose, 06108 Nice Cedex 2, France \\
e-mail: sornette@moho.ess.ucla.edu\\
phone: (310) 825 2863~~~~~fax: (310) 206 3051\\
Anders Johansen is a research assistant professor\\
Didier Sornette is a director of research at CNRS, France
and a professor of Geophysics at UCLA}

\date{\today}
\maketitle

\pagebreak

\begin{center}
\centering{Large Stock Market Price Drawdowns Are Outliers}
\end{center}

\abstract{Drawdowns (loss from the last local maximum to the next local minimum) offer
a more natural measure of real market risks than the variance, the value-at-risk
or other measures based on fixed time scale distributions of returns. Here, we
extend considerably our previous analysis of drawdowns by analyzing the major financial
indices, the major currencies, gold, the twenty 
largest U.S. companies in terms of capitalisation as well as nine others
chosen randomly. Approximately $98\%$ of the distributions of drawdowns
is well-represented by an exponential (or a minor
modification of it with a slightly fatter tail), while the largest to the few ten
largest drawdowns are occurring with a significantly larger rate
than predicted by the exponential. This is confirmed by 
extensive testing on surrogate data. Very large drawdowns thus
belong to a different class of their own and call for a specific amplification
mechanism. Drawups (gain from the last local minimum to the next local
maximum) exhibit a similar behavior in only about half the markets examined.
}

\thispagestyle{empty}
\pagenumbering{arabic}
\newpage
\setcounter{page}{1}

\section{Introduction}

It is now quite universally accepted that the distribution of asset
returns is not only leptokurtotic but belongs to the class of fat tailed
distributions. More formally, it has been shown that the tails of the
distribution of returns follow approximately a power law
$P({\rm return}>x) \sim C/x^{\mu}$,
with estimates of the 'tail index' $\mu$ falling in the range 2 to 4 
[de Vries (1994); Lux (1996); Pagan (1996); Guillaume et al. (1997); Gopikrishnan et al.
(1998)].  This
implies that the second and probably the third moment of the distribution are
finite. Extrapolating this distribution to infinite values,
the fourth and higher moments are predicted to be mathematically infinite.
This approximate law seems to hold for returns calculated over time scales
ranging from a few minutes to about three weeks while the distributions are consistent
with a slow convergence to Gaussian behavior at larger time scales
[Gopikrishnan et al. (1998); Plerou et al. (1999)].
An alternative description with finite moments of all orders but still with
fat tails
has been suggested in terms of stretched exponential, also known as
sub-exponential or Weibull (with exponent less than one) distributions
[Laherr\`ere and Sornette (1998)] (see [Sornette (2000)] for a
synthesis of maximum likelihood estimators for the Weibull distributions).

This ``one-point'' statistics is however far from sufficient for characterizing market
moves [Campbell et al. (1997); Lo and MacKinlay (1999)]. 
Two-point statistics, such as correlations of price returns and of 
volatilities (with the persistence phenomenon
modeled by ARCH processes and its generalizations)
offers important and complementary but still limited informations. 
In principle, one would like to have access to the full hierarchy of 
multiple-point correlations functions, but this is not attainable in 
practice due to finite statistics. A short-cut is to realize that 
using a fixed time scale, such as daily returns, is not adapted
to the real dynamics of price moves and that relatively low-order
statistics with suitable adjustments to the relevant time scales of the market
may be more efficient descriptions.
Indeed, physical time is probably not the
proper quantity to characterize the flow of information and the rhythm of trading. 
Clark (1973) first noticed that subordinated processes, in which time is itself a
stochastic process, provide a natural mechanism for fat tails resulting
from the fact that the distribution of increments of
subordinate random walks is a mixture of normal distributions (which is
generically leptokurtic). A possible candidate for the stochastic time process
(the subordinator) is the transaction volume [Clark (1973)] or the number of trades 
[Geman and An\'e (1996)]. These processes
can be generalized into multifractal subordinated processes [Mandelbrot (1997)]. 
In a similar vein, M\"uller et al. (1995), Guillaume et al. (1995) and Dacorogna
et al. (1996)
have advocated the concept of an elastic time that expands periods of high
volatility and contracts those of low volatility, thus capturing better the relative
importance of events on the market. Related empirical works have shown that
return volatilities exhibit long-range correlations
organized in a hierarchical way, from large time scales to small time scales
[Ghashghaie et al. (1996); M\"uller et al. (1997);
Dacorogna et al. (1998); Arneodo et al. (1998); Ballocchi et al. (1999); Muzy et al. (2000, 2001);
Breymann et al. (2000)]. 
All these approaches suggest that
a fixed time scale is not adequate for capturing the perception of
risk and return experienced by traders and investors. There is thus a large
potential gain in being time-adaptive rather than rigid in order to gain insight 
into the dynamics of financial markets.

Extreme value theory (EVT) provides an alternative approach, still based on the 
distribution of returns estimated at a fixed time scale. Its most practical 
implementation is based on the so-called ``peak-over-threshold'' distributions
[Embrechts et al., 1997; Bassi et al., 1998], which is founded
on a limit theorem known as the Gnedenko-Pickands-Balkema-de Haan theorem which 
gives a natural limit law for peak-over-threshold values in the form of the 
Generalized Pareto Distribution (GPD), a family of distributions with two parameters
based on the Gumbel, Weibull and Frechet extreme value distributions. 
The GPD is either an exponential or has a power law tail. Peak-over-threshold
distributions put the emphasis on the characterization of the tails of 
distribution of returns and have thus been scrutinized for their potential
for risk assessment and management of large and extreme events
(see for instance [Phoa, 1999; McNeil, 1999]). In particular,
extreme value theory provides a general foundation for the estimation of 
the value-at-risk for very low-probability ``extreme'' events. There
are however severe pitfalls [Diebold et al., 2001] in the use of extreme
value distributions for risk management because of its reliance on
the (unstable) estimation of tail probabilities. In addition, the EVT
literature assumes independent returns, 
which implies that the degree of fatness in the tails decreases as 
the holding horizon lengthens (for the values
of the exponents found empirically).  Here, we show that this is not the case:
returns exhibit strong correlations at special times precisely 
characterized by the occurrence of extreme events, the regime that EVT aims
to describe. This suggests to re-examine EVT and extend it to variable
time scales, for instance by analyzing the EVT of the distribution of 
drawdowns and drawups. 

In order to address the problem of the possible effect of correlations 
that may occur at different time scales, we focus here on the statistics of drawdowns
(and their complement, the drawups). We define a drawdown as a
persistent decrease in the price (specifically the
closing price) over consecutive days. A drawdown is thus
the cumulative loss from the last maximum to the next minimum of the price.
Symmetrically, a drawup is defined as the change (in percent) between a
local minimum and the following maximum, {\it i.e.}, a drawup is the
event that follows the drawdown and {\it vice versa}. A drawup is
a drawdown for an agent with a short (sell) position on the corresponding
market.

Drawdowns are highly relevant: they measure directly the cumulative loss that an investment
can incur. They also quantify the worst case scenario of an investor buying at the
local high and selling at the next minimum. It is thus worthwhile to ask 
if there is any structure in the distribution
of drawdowns. Notice that drawdowns embody a subtle dependence since they
are constructed from runs of the same sign variations. They are not defined
over a fixed time scale. Some drawdowns will last only one day, other five days
or more.  Their distribution measures how successive
drops can influence each other and construct a persistent process, not
captured by the distribution of returns nor by the
two-point correlation function. In Appendix A, 
we show that the distribution of drawdowns for independent
price increments $x$ is asymptotically an exponential 
(while the body of the distribution is Gaussian [Mood (1940)]) when the distribution
of $x$ does not decay more slowly than an exponential, {\it i.e.}, belong to the class
of exponential or super-exponential distributions. In contrast, for sub-exponentials
(such as stable L\'evy laws, power laws and stretched exponentials), the 
tail of the distribution of drawdowns is asymptotically the same as the 
distribution of the individual price variations. 
Since stretched exponentials have been found to offer 
an accurate quantification of price variations 
[Lah\'errere and Sornette, 1998; Sornette et al., 2000; Andersen and Sornette, 2001]
thus capturing a possible sub-exponential behavior
and since they contain the exponential law as a special case (exponent $z=1$ in
the definition (\ref{stretched})), we
shall take the stretched exponential law as our null hypothesis.

Our emphasis on drawdowns is motivated by two considerations: 1) 
drawdowns are important measures of risks used by practitioners because they represent 
their cumulative loss since the last estimation of their wealth. It
is indeed a common psychological trait of people to estimate a loss
by comparison with the latest maximum wealth; 2) drawdowns automatically capture an
important part of the time dependence of price returns, similarly to the 
so-called run-statistics often used in statistical testing [Knuth (1969)]
and econometrics [Campbell et al. (1997); Barber and Lyon (1997)].
As we have previously showed [Johansen and Sornette (1998,2000b)], 
the distribution of drawdowns 
contains an information which is quite 
different from the distribution of returns over a fixed time scale. 
In particular, a drawndown embodies the interplay 
between a series of losses and hence measures a ``memory''  of the 
market. Drawdowns will examplify the effect of correlations in price 
variations when they appear, which must be taken into account for a 
correct characterisation of market price variations. They are direct
measures of a possible amplification or ``flight of fear'' where previous 
losses lead to further selling, strengthening the downward trend,
occasionally ending in a crash. We stress
that drawdowns, by the ``elastic'' time-scale used to define them,
are effectively function of several higher order correlations at the same time.

The data used in our work comprises 
\begin{enumerate}
\item major world financial indices: the Dow Jones, Standard \& Poor, Nasdaq
Composite, TSE 300 Composite (Toronto, Canada), All Ordinaries
(Sydney stock exchange, Australia), Strait Times (Singapore stock exchange), 
Hang Seng (Hong Kong stock exchange), Nikkei 225 (Tokyo stock exchange, Japan), 
FTSE 100 (London stock exchange, U.K.), CAC40 (Paris stock exchange, France), 
DAX (Frankfurt stock exchange, Germany), MIBTel (Milan stock exchange, Italy);
\item currencies: US\$/DM, US\$/Yen, US\$/CHF;
\item gold;
\item the $20$ largest companies in the US market in terms of capitalisation, as 
well as 9 others taken randomly in the list of the 50 largest companies 
(Coca Cola, Qualcomm, Appl. Materials, Procter\&Gamble, JDS Uniphase, General
Motors, Am. Home. Prod., Medtronic and Ford).
\end{enumerate}
These different data have not the same time span, largely due to different lifespan,
especially for some recent ``new technology'' companies. 
In this selection of time series, we are far from exhaustive but have a 
reasonable sample for our purpose: as we shall see, with the exception of 
the index CAC40 (the ``French exception''?), all time series 1-3 exhibit clear
outlier drawdowns. This suggests that outliers constitute an ubiquitous feature of 
stock markets, independently of their nature.

In the next section 2, we discuss some properties of drawdowns.
In particular, we analyze the drawdowns of a
model with no correlation but strong dependence and present preliminary 
evidence of intermittent dependence on the Dow Jones Industrial Average index.

In section 3, we present our results for the drawdowns and the drawups of the 
major world financial markets, of the major world currencies and of gold. 
While it is well-known that price returns are essentially
uncorrelated, section 3 will show that strong correlations do 
appear at special times when large drawdowns and drawups occur: the distribution 
of large drawdowns and large drawups is strongly non-exponential with a much fatter tail.
Since this anomalous behavior is observed only for the largest
drawdowns and drawups, in some cases up to a few tens of the
largest events, these very large drawdowns and drawups can thus be considered to be
{\it outliers} because they do not conform to the model suggested by the 
main part of the data. This points to the brief appearence of a
dependence in successive drops leading to an amplification which makes these
drops special. The results confirm and extend our previous 
announcements [Johansen and Sornette (1998,2000b)]. We test these results by
constructing error plots and by adding noise to investigate the robustness
of the distribution of drawdowns and drawups.

Section 4 presents the evidence that large drawdowns and drawups are outliers 
for 29 of the largest US companies.
The distribution of 
drawups is found significantly different from that of drawdowns. While drawups of amplitude
larger than $15\%$ occur about twice as often as drawdowns of the same amplitude,
the case for the largest drawups to be outliers is less clear-cut. Half the 
time series have their largest drawups significantly larger than explained by
the bulk of the distribution, the converse is observed for the other half. It
thus seems that outlier drawups is a less conspicuous feature of financial series
with more industry specificities than the ubiquitous outlier drawdowns observed
essentially in all markets.

Section 5 presents statistical tests using surrogate data that confirm that large
drawdowns are outliers with a large degree of significance.

Section 6 summarizes our results and concludes with a discussion on the implications
for risk management.

Appendix A derives the distribution of drawdowns for a large class of distributions
of returns in the restricted case of independent returns. The results
justify our choice of the exponential and stretched exponential null
hypothesis. Appendix B gives the
confidence interval for drawdowns when the distribution of returns is in the
exponential or superexponential class.

\section{Some properties of drawdowns}

\subsection{A model with no correlation but strong dependence} 

To see how subtle dependences in successive price variations
are measured by drawdowns, consider the simple but illustrative
toy model in which the price increments $\delta p(t)$ are given by
[Robinson (1979); Hsieh (1989)]
\be
\delta p(t) = \epsilon(t) + \epsilon(t-1) \epsilon(t-2)~,   \label{nvnvnv}
\ee
where $\epsilon(t)$ is a white noise process with zero mean and unit
variance.  Then, the expectation ${\rm E}(\delta p(t))$ as well as 
the two-point correlation ${\rm E}(\delta p(t) \delta p(t'))$ for $t \neq t'$
are zero and $\delta p(t)$ is also a white noise process. However, the 
three-point correlation function ${\rm E}(\delta p(t-2) \delta p(t-1)\delta p(t))$
is non-zero and equal to $1$ and the expectation of $\delta p(t)$
conditioned on the knowledge of the {\it two} previous increments
$\delta p(t-2)$ and $\delta p(t-1)$ is non-zero and equal to 
\be
{\rm E}(\delta p(t)|\delta p(t-2),\delta p(t-1)) = \delta p(t-2) \delta p(t-1)~,
\ee
showing a clear predictive power. This leads to a very distinct signature in the distribution
of drawdowns. To simplify the analysis to the extreme and make the 
message very clear, let us restrict to the case
where $\epsilon(t)$ can only take two values $\pm 1$, Then, $\delta p(t)$ can
take only three values $0$ and $\pm 2$ with the correspondance
\ba
\epsilon(t-2), \epsilon(t-1), \epsilon(t) ~~~~~&\to& ~~~~~\delta p(t)\\
+++ ~~~~~&\to&  ~~~~~+2   \nonumber \\
++- ~~~~~&\to&  ~~~~~0 \nonumber\\
+-+ ~~~~~&\to&  ~~~~~0 \nonumber\\
+-- ~~~~~&\to&  ~~~~~-2 \nonumber\\
-++ ~~~~~&\to&  ~~~~~0 \nonumber\\
-+- ~~~~~&\to&  ~~~~~-2 \nonumber\\
--+ ~~~~~&\to&  ~~~~~+2 \nonumber\\
--- ~~~~~&\to&  ~~~~~0  \nonumber
\ea
where the left colum gives the three consecutive values $\epsilon(t-2), \epsilon(t-1),
\epsilon(t)$ and the right column is the corresponding price increment $\delta p(t)$.
We see directly by this explicit construction that $\delta p(t)$ is a white noise
process.  However, there is a clear predictability and the distribution of drawdowns
reflects it: there are no drawdowns of duration larger than two time steps. 
Indeed, the worst possible drawdown corresponds to the following sequence for
$\epsilon$: $--+--$. This
corresponds to the sequence of price increments $+2, -2, -2$,
 which is either stopped by a $+2$ if the next 
$\epsilon$ is $+$ or by a sequence of $0$'s interupted by a $+2$ at the first
$\epsilon = +$. While the drawdowns of the process $\epsilon(t)$ can in principle be 
of infinite duration, the drawdowns of $p(t)$ cannot. This shows that
the structure of the process $\delta p(t)$ defined by (\ref{nvnvnv})
has a dramatic signature in the distribution of drawdowns in $p(t)$ and 
illustrates that drawdowns, rather
than daily or weekly returns or any other fixed time scale
returns, are useful time-elastic measures of price moves.

\subsection{Variations on drawdowns}

There is not a unique definition of 
drawdowns as there are several possible choices for the 
peak where the drawdown starts from.
For instance, the ``current drawdown'' is the
amount in percent that a portfolio has declined from its most recent peak. 
The maximum drawdown is the largest amount (in percent) that a fund or
portfolio dropped from a peak over its lifetime. Grossman and Zhou (1993)
analyze the optimal risky investment for an investor not willing to lose
at each point in time a drawdown (calculated relative to the highest value of the
asset in the past) larger than a fixed percentage of the 
maximum value his wealth has achieved up to that time. 
Maslov and Zhang (1999)
have investigated the distribution of ``drawdowns from the maximum'',
where the drawdowns are also calculated relative to the  highest value of the
asset in the past. This distribution is defined only for an asset exhibiting 
a long-term upward trend and can be shown, for uncorrelated returns, to 
lead to power law distributions. In this paper, drawdowns will be defined
as the decrease in percent from a local maximum to the following local 
minimum after which the price again increases.

\subsection{Preliminary evidence of intermittent dependence}

Appendix A shows that the distribution of drawdowns for independent
price increments $x$ is asymptotically an exponential when the distribution
of $x$ does not decay more slowly than an exponential, {\it i.e.}, belong to the class
of exponential or super-exponential distributions. In contrast, for sub-exponentials
(such as stable L\'evy laws, power laws and stretched exponentials), the 
tail of the distribution of drawdowns is found to be asymptotically the same as the 
distribution of the individual price variations.

These results hold as long as the assumption, that
successive price variations are uncorrelated, is a good approximation. 
There is a large body of 
evidence for the correctness of this assumption for the largest fraction 
of trading days [Campbell et al. (1997)]. However, consider, for instance, 
the 14 largest drawdowns that has occurred in the Dow Jones Industrial 
Average in the last century. Their characteristics are presented in table
\ref{largedddj}. Only 3 lasted one or two days, whereas 9 lasted four
days or more. 
Let us examine in particular the largest drawdown. It started on Oct. 
14, 1987 (1987.786 in decimal years), lasted four days and led to a 
total loss of $-30.7\%$. This crash is thus a run of four consecutive 
losses: first day the index is down with $3.8\%$, second day with $6.1\%$, 
third day with $10.4\%$ and fourth with $30.7\%$. In terms of consecutive 
losses this correspond to the following sequence of 
daily losses: $3.8\%$, $2.4\%$, $4.6\%$ and $22.6\%$ on 
what is known as the Black Monday of 19th Oct. 1987.

The observation of large successive drops is suggestive of the existence
of a transient correlation. To make this point clear, consider an hypothetical 
drawdown of $30\%$ made up of three successive drops of $10\%$ each.
A daily loss of $10\%$, while severe, is not that uncommon and occurs
typically once every four years (or one in about $1000$ trading days).
It thus corresponds to a probability of $10^{-3}$. Assuming that the
three successive drops of $10\%$ are uncorrelated, the probability of 
the drawdown of $30\%$ is $[10^{-3}]^3=10^{-9}$, {\it i.e.}, corresponds to one
event in four millions years. In other words,
the lack of correlations is completely in contradiction with empirical
observations.

For the Dow Jones, this reasoning can be adapted as follows. 
We use a simple functional form for the distribution of daily losses, 
namely an exponential distribution with decay rate $1/0.63\%$ obtained
by a least-square fit. This quality of the exponential model
is confirmed by the direct calculation of the average loss amplitude equal
to $0.67\%$ and of its standard deviation equal to $0.61\%$ (recall
that an exact exponential would give the three values exactly equal).
Using these numerical values, the probability for a drop equal to or
larger than $3.8\%$ is $\exp(-3.8/0.63)= 2.4~10^{-3}$ (an event occurring
about once every two years);  the probability for a drop equal to or
larger than $2.4\%$ is $\exp(-2.4/0.63)= 2.2~10^{-2}$ (an event occurring
about once every two months);  the probability for a drop equal to or
larger than $4.6\%$ is $\exp(-4.6/0.63)= 6.7~10^{-4}$ (an event occurring
about once every six years). Under the hypothesis that these three first events are
uncorrelated, they together correspond to a probability of occurrence of 
$3.5 ~10^{-8}$, {\it i.e.}, one event in about $11,000$ years, an extremely rare
possibility. It turned out that, not only this exceedingly rare drawdown occurred
in Oct. 1987 but, it was followed by an even rarer event, a single day drop of $22.6\%$.
The probability for such a drop equal to or
larger than $22.6\%$ is $\exp(-22.6/0.63)= 2.6~10^{-16}$ (an event occurring
about once every $10^{14}$ years). All together, under the hypothesis that 
daily losses are uncorrelated from 
one day to the next, the sequence of four drops making the largest drawdown
occurs with a probability $10^{-23}$, {\it i.e.}, once in about $4$ thousands
of billions of billions years. 
This clearly suggests that the hypothesis of uncorrelated daily 
returns is to be rejected and that drawdowns 
and especially the large ones may exhibit intermittent correlations 
in the asset price time series.

\section{Cumulative distributions of drawdowns and drawups for major world indices, 
currencies and gold}

In this section, we present three pieces of evidence supporting the existence
of two classes of drawdowns (and to a lesser degree of drawups). 

First for all studied assets, we construct on the same graphs the
cumulative distribution of drawdowns and the complementary cumulative
distribution of drawups. Each of these cumulative distributions is fitted to 
the null hypothesis taken as an exponential or a stretched exponential
represented in the graphs by the continuous lines.
The cumulative stretched distribution is defined by
\be \label{stretched}
N_c\lp x\rp = A ~\exp \left( - (|x|/\chi)^z \right) ~,
\ee
where $x$ is either a drawdown or a drawup. 
In order to stabilize the fit, it has been performed 
as $\log(N_c) = \log(A) - B|x|^z$, where $A$ is the total number 
of drawdowns (or drawups) and hence is fixed \footnote{This is equivalent 
to a normalisation of the corresponding probability distribution.}. 
The characteristic scale $\chi$ is related to the
coefficient $B$ by the relation $\chi= 1/B^{1/z}$.
When $z<1$ (resp. $z>1$), $N_c\lp x\rp$ is a stretched exponential
or sub-exponential (resp. super-exponential). The special case $z=1$
corresponds to a pure exponential. In this case, $\chi$ is nothing but
the standard deviation of $|x|$.

The success in using the stretched exponential for
parameterising the data does not of course mean that better 
parametrizations cannot be found, neither does it prove that the processes
governing the dynamics of the main part of the distribution is
giving stretched exponential distributions. As we said above, its virtue
lies in the fact that it
is a straightforward generalisation of the pure exponential, which 
represents the simplest and natural null-hypothesis for uncorrelated price variations.
We find that the stretched exponential model 
parameterizes $\approx 98\%$ of all data points, while the $2\%$ largest values
are the ``outliers''. We find that twelve markets have a value of the exponent $z$
below $1$ and only five have a value larger than $1$.

The second piece of evidence is provided by plotting
the relative error between the empirical cumulative
distributions and the stretched exponential fits. This graphical 
presentation makes very apparent the transition between the two regimes.

Third, we test for the robustness of these results by adding noise 
of different amplitudes to the data.

\subsection{Major financial indices}

Figures \ref{djdd-dunorm} to \ref{nasdd-dunorm}
show the cumulative distributions $N_c\lp x\rp$ 
of drawdowns and complementary distributions of drawups
in three major US indices: The Dow Jones 
Industrial Average, the S\&P500 and the Nasdaq Composite. 
The two continuous lines show the fits of these
two distributions with the stretched exponential distribution defined by
formula (\protect\ref{stretched}). 
Up to drawdawns of amplitudes $\approx 10\%$, the stretched exponential
distribution provides a faithful description while above $\approx 10\%$,
the distribution ``breaks away'', most abruptly for the \dj \ and the S\&P500. 

Both the Dow Jones Industrial Average and the 
Nasdaq Composite indice  show a distinct upward curvature in the
distribution of drawdowns corresponding to $z\approx 0.82\pm 0.02$, 
whereas the S\&P500 is closer to a pure exponential with $z\approx 0.90$, 
see table \ref{ddindex}. Note that an exponent $z < 1$ corresponds to 
a fatter tail than an exponential. The law (\ref{stretched}) is also known
in the mathematical literature as a ``subexponential'' and in the engineering
field as a stretched distribution (see chap.~6 in (Sornette, 2000) for useful
informations on this family of distributions and how to calibrate them). Our aim is not
to defend the model (\ref{stretched}) but rather to use it as a convenient 
and parsimonious tool. This family (\ref{stretched}) 
is particularly adapted to our problem because 
it both contains the null hypothesis of an
 exponential distribution $z=1$ valid for exponential and superexponential
 distributions and also allows us to describe subexponential distributions
 (see Appendix A). Allowing $z$ to depart from
 $1$ then provides us a simple and robust measure of the deviation from an exponential. 
 We actually find that allowing for an even fatter tail is not enough to 
 account for the very largest drawdowns which wildly depart from this description. 
 Playing the devil's advocate, one could argue that the strong departure from 
 model (\ref{stretched}) of the few
 largest drawdowns observed in figures \ref{djdd-dunorm} to \ref{nasdd-dunorm}
 is not a proof
 of the existence of outliers, only of the inadequacy of the model (\ref{stretched}).
 It is true that we can never prove the existence of outliers in an absolute sense.
 However, we have calculated a reasonable null-hypothesis, provided 
 a family of distributions that extends the null-hypothesis to account for larger moves. 
 Nevertheless, we find that the largest drawdowns are utterly different
from the distribution containing approximately $98\%$ of the data. 
We find this to be an important message which may
lead to a better understanding of how the financial markets react in
times of large losses. This has been tested within a rigorous 
statistical procedure in [Sornette and Johansen, 2001].

There is a clear asymmetry between drawdowns and drawups. The asymmetry
occurs more clearly in the bulk of the distribution for the Dow Jones Industrial Average
where $z=0.84 \pm 0.01$ for drawdowns compared to $z=0.99 \pm 0.01$ for drawups. 
In the case of the \dj, the change of regime to extreme events is maybe
even stronger for drawups than for drawdowns. The difference is
weaker for the S\&P500 index with $z=0.90 \pm 0.01$ for drawdowns compared to 
$z=1.03 \pm 0.02$ for drawups. The change of regime occurs clearly
for both drawdowns and drawups with the difference that a very strong
outlier exists only for drawdowns. The asymmetry in the bulk
of the distributions of drawdowns and drawups is also apparent for the
Nasdaq composite index with $z=0.80 \pm 0.02$ for drawdowns compared to 
$z=0.90 \pm 0.02$ for drawups. However, the asymmetry is much more pronounced
in the extreme event regime: no clear change of regime is observed for drawups
whose full distribution is satisfactory described by the stretched
exponential distribution (\ref{stretched}) while a strong departure
from (\ref{stretched}) is observed for the five large drawdowns.

Is this behavior confined to U.S. markets or 
is it a more general feature of stock market behavior? In order to answer
this question, we have analysed the main stock market index of the remaining 
six G7-countries as well as that of Australia, Hong-Kong and Singapore. The results of
this analysis is shown in figures \ref{candd-dunorm} to \ref{itdd-dunorm} and more
quantitatively in table \ref{ddindex}. Quite remarkably, for 
the drawdowns, we find that all
markets except the French market, 
the Japanese market being on the borderline\footnote{An explanation 
for this is that the Japanese stock market exhibited a general decline
from 1990 to early 1999 which is more than a third of the data set. The total
decline was approximately two thirds in amplitude.}, show the same qualitative behavior 
exhibiting a change of regime and the presence of ``outliers''. The
Paris stock exchange is the only exception as the distribution of drawdowns 
is an almost perfect exponential. It may be that the observation time
used for CAC40 is not large enough for an outlier to have occurred. If we
compare with the Strait Times index (Singapore stock exchange)
shown in figure \ref{singdd-dunorm}, we see that all the distribution
except the single largest drawdown is very well-fitted by a
stretched exponential. The presence
or absence of this outlier thus makes all the difference.

The results obtained
for the drawups are not as clear-cut as for the drawdowns. 
Whereas the distributions for the \dj \ and the S\&P500 
clearly exhibits a similar behavior to that of the drawdowns, this is 
not the case for the TSE 300, the Strait times, the Hang Seng, 
the FTSE 100, the DAX and the MIBTel indices. The other 
markets are at the borderline of significance\footnote{The reader 
is again reminded of the Japanese recession of 1990-1999.}. 
In the one case where we have sufficient statistics
(for the \dj), we have successfully parameterized the tail of the drawup 
distribution with another stretched exponential (not shown), see table \ref{duindex}. 
However, how to interpret the interpolation of such two rather excellent fits 
is difficult. With respect to the other markets that clearly exhibit 
outliers, it is impossible to give a qualified estimation of the distribution 
of the outliers due to the limited statistics just as with the drawdowns.

To quantify the statistical significance of the fluctuations obtained
in the exponent $z$ reported in table \ref{ddindex}, we use
the asymptotic covariance matrix of the 
estimated parameters, the typical scale $\chi \equiv 1/B^{1/z}$ and 
the exponent $z$ calculated by Thoman et al. (1970):
\ba
{\sigma_{\chi} \over \chi} &=& {1.053 \over z}~ {1 \over \sqrt{N}}~,  \label{nbvbvnx}\\
{\sigma_{z} \over z} &=& {0.78 \over \sqrt{N}}~, \label{nvbklkz}
\ea
where $N$ is the sample size (number of drawdowns (resp. drawups) in our data sets) and
$\sigma_{\chi}$ (resp. $\sigma_{z}$) is the standard deviation of $\chi$ (resp. $z$).
The fluctuations in the exponent $z$ are statistically significant: 
excluding the CAC40, we find an average $z=0.87$ with a standard
deviation of $0.04$ which is about two times larger than expected from the 
statistics (\ref{nvbklkz}). These two expressions (\ref{nbvbvnx}) 
and (\ref{nvbklkz}) are used to give the one-sigma error bars in the tables.

\subsection{Gold and in currencies \label{sectiongold}}

To test whether the existence of outliers beyond a stretched exponential regime
apply more generally, we now analyze currencies and gold.
As seen in figure \ref{usdmdd-dunorm} for the US\$/DM exchange rate, 
figure \ref{usyendd-dunorm} for the US\$/YEN exchange rate, 
figure \ref{uschfdd-dunorm} for the US\$/CHF exchange rate and 
figure \ref{audd-dunorm} for gold, a similar behavior is observed. The exponent $z$
for drawdowns tends to be larger with a smaller
characteristic drawdown $\chi$ (thinner tails) for currencies compared to the indices.
The gold market and the currencies confirm the previous evidence by
providing us with some of the strongest cases for the presence
of outliers, together with the US 
indices, TSE 300, All Ordinaries, FTSE 100 and the DAX indices.

The bulk of the distributions of drawdowns and drawups are well-fitted
by the stretched exponential model with an almost perfect symmetry 
except for the US\$/YEN exchange rate. Indeed, the exponents $z$ and
characteric scales $\chi$ are basically identical for drawdowns and 
drawups, except for the US\$/YEN exchange rate which exhibits
a strong asymmetry in the parameter
$\chi=1.17 \pm 0.03$ for drawdowns and $\chi=0.79 \pm 0.02$ for drawups, while
$z=0.90 \pm 0.02$ is the same for both drawdowns and drawups.
Interestingly, the asymmetry is further strengthened by the fact that 
the ``outlier'' regime is only observed for drawdowns while all the drawups
are very well-described by the stretched exponential distribution, as
seen in figure \ref{usyendd-dunorm}.

In general, it is expected that, for most currency pairs, the pdf for the return is symmetric.
We note however that this is not always true: Laherr\`ere and Sornette [1998] documented
that the distribution of returns of the French franc expressed in German mark (before
the introduction of the Euro) is asymmetric
with a fatter tail for the losses than for the gains. Thus, the slow and continuous depreciation
of the French franc against the German mark over the years preceding
the unique European currency has occurred by a 
globally fatter distribution of losses rather than by a slow drift superimposed upon
a symmetric distribution. It seems that a similar asymmetry of the daily return 
distribution of the US \$ expressed in Yen may explain in part the asymmetry of the drawdowns:
from 1972 to 1999, the US \$ has lost a factor $3$ in its value compared to the Yen.
As for the French franc/German mark case, this loss has occurred by a slighly fatter
tail of the distribution of losses compared to the distribution of
gains of the US \$ in Yen. The construction of drawdowns seems to enhance this 
asymmetry of the distribution of daily returns and indicates 
that the slow depreciation of the US \$ has occurred in bursts
rather as a continuous drift. 
To investigate further this question, we analyzed the the GPB/US\$ exchange rate
and the GPB/DM distributions. We find that they are to a good approximation 
described by a pure exponential behaviour with no outliers.
This may be the
evidence of a singular increasing role played by the Japanese Yen over the 
thirty years since the early 1970s.

\subsection{Error plots}

Figures \ref{alldderror} and \ref{alluperror} show
the differences (error plots) between the cumulative 
distributions of drawdowns (respectively drawups) and the best fits 
with the stretched exponential
model (\ref{stretched}) for the DJIA, the S\&P500, the Nasdaq composite index, 
the German mark in US \$ and Gold. 
These two figures clearly identify two regimes.
\begin{enumerate}
\item The bulk of the distributions correspond to errors fluctuating
around $0$. This characterizes typically the drawdowns and drawups less than 
about $5-7\%$ representing about $98\%$ of the data points. 
This is nothing but a restatement of the good fit provided
by the stretched exponential model. 
\item In contrast, for drawdowns and drawups larger than 
$5-7\%$, one observes a striking systematic deviation characterizing the 
``outlier'' regime.
\end{enumerate}

\subsection{Tests of robustness by adding noise \label{hghngw}}

As we already mentioned, the definition of drawdowns
and drawups may vary slightly under small changes in the data. For instance,
consider a run of daily losses in which one day witnessed a small loss of $-0.05\%$.
Had the market changed this value to a small gain of $+0.05\%$, this drawdown would have
been replaced by two drawdowns separated by one slightly positive day. 
Since one should not expect such days with very small
positive or negative returns to play a role in the significance of the cumulative
losses and gains captured by the drawdowns, the question therefore arises
whether our results are robust with respect to the addition of a noise that may modify these 
small returns. 

In order to address this question, we 
define the ``time series neighborhood'' of a given price trajectory as follows.
\begin{enumerate}
\item We define the $i$-th return $r(i)= \ln[{\rm price}(i)]-\ln[{\rm price}(i-1)]$.

\item We perform the transformation $r(i) \to r_A(i) = r(i) + A ~{\rm Ran} ~\sigma$
where ${\rm Ran}$ is a i.i.d. random noise with uniform distribution in $[-1, +1]$
and $\sigma$ is the standard deviation of the returns. We have performed
tests with $A=0$ (no added noise), $A=0.5, 1$ and $2$.

\item We exponentiate the time series $r_A(i)$ to get a new noisy price
time series 
\be
p_A(i) = p_0 \exp \left[r_A(i)+r_A(i-1) +r_A(i-2)+...+r_A(1)\right]~.
\ee

\item We then construct the runs of successive losses (drawdowns) and of 
successive gains (drawups) for this noise price time series  $p_A(i) $.
\end{enumerate}

If the anomalous behavior of the drawdowns in the ``outlier'' regime
is due to high order correlations of the return, it should not be
destroyed by the addition of noise and
the distribution of outliers should be similar to the one computed with 
the initial price series. This is indeed verified in figures 
\ref{djnoiseline}-\ref{nasnoiseline} which show
the cumulative distribution of drawdowns
and complementary distributions of drawups for the
 ``time series neighborhoods'' of the DJIA, the S\&P500 and the
 Nasdaq composite index for the different values of $A=0$, $A=0.5, 1$ and $2$.
 
These tests are encouraging. Other tests that will be reported elsewhere
[Johansen and Sornette, in preparation]
confirm this picture. They involve
either ignoring increases (decreases)
of a certain fixed magnitude (absolute or relative to the price) or 
ignorings increases (decreases) over a fixed time horizon and
in both cases letting 
the drawdown (drawup) continue. Both approaches provide a natural ``coarse-graining'' 
of the drawdowns and drawups and confirm/strengthen the validity of our results.

\section{Drawdowns and drawups in stock market prices of major US companies}

\subsection{``Universal'' normalized distributions}

We now extend our analysis to the very largest companies
in the U.S.A. in terms of capitalisation (market value). The ranking is 
that of Forbes at the beginning of the year 2000. We have chosen the top 
20, and in addition a random sample of other companies, namely
number 25 (Coca Cola), number 30 (Oualcomm), number 35 (Appl. Materials), 
number 39 (JDS Uniphase), number 46 (Am. Home Prod.)
and number 50 (Medtronic). 
Three more companies have been added in order to get
longer time series as well as representatives for the automobile sector. 
These are Procter \& Gamble (number 38), General Motors (number 43) and 
Ford (number 64). This represent a non-biased selection based on objective
criteria. 

To avoid a tedious repetition of many figures, we group the cumulative
distributions of drawdowns and complementary cumulative distributions
of all these stocks in the same figure \ref{rescaleall-1}. In order to 
construct this figure, we have fitted the stretched
exponential model (\ref{stretched}) to each distribution and obtained the corresponding parameters
$A$, $\chi$ and $z$ given in tables \ref{ddtable} and \ref{dutable}. 
We then construct the
normalized distributions 
\be
N_C^{(n)}(x) = N_c\lp (|x|/\chi)^z \rp /A
\label{ghbwhsz}
\ee
using the triplet $A$, $\chi$ and $z$ which is specific to each distribution.
Figure \ref{rescaleall-1} plots the expression (\ref{ghbwhsz}) for each
distribution, {\it i.e.}, $N_c/A$ as a function of 
$y \equiv {\rm sign}(x)~(x/\chi)^z$. If the stretched
exponential model (\ref{stretched}) held true for all the drawdowns and all the drawups,
all the normalized distributions should collapse exactly onto the 
``universal'' functions
$e^y$ for the drawdowns and $e^{-y}$ for the drawups. We observe that this is
the case for values of $|y|$ up to about $5$, {\it i.e.}, up to typically $5$ standard
deviations (since most exponents $z$ are close to $1$ as seen
in tables \ref{ddtable} and \ref{dutable}), beyond which there is a clear upward
departure observed both for drawdowns and for drawups. Comparing with the
extrapolation of the normalized stretched exponential model $e^{-|y|}$,
the empirical normalized distributions give about $10$ times too many drawdowns
and drawups larger than $|y|=10$ standard deviations and more the $10^4$
too many drawdowns and drawups larger than $|y|=20$ standard deviations.
Note that for AT\&T, a crash of $\approx 73\%$ occurred which lies
beyond the range shown in figure \ref{rescaleall-1}.

\subsection{Detailed results on the distributions of drawdowns and drawups}

A detailed analysis of each individual price distribution of drawdowns shows that 
the five largest companies (MicroSoft, Cisco, General Electric, Intel 
and Exxon-Mobil) clearly exhibits the same features as those for the
major financial markets. Of the remaining 24, for all but America Online and JDS
Uniphase, we find clear outliers but also a variety of different tails of
the distributions. The main difference is
in the value of the exponent $z$, which is $\approx 1$ or larger. This means 
that the distributions tends to bend downward instead of upward, thus 
{\it emphasising} the appearance the outliers. It is interesting to note
that the two companies, America Online and JDS Uniphase, whose distributions
did not exhibit outliers are also the two companies with by far the largest number per year
of drawdowns of amplitude above $15\%$ (close to $4$), see table 
\ref{ddtable}.

For the drawups in the price of major US companies, the picture
is less striking when compared to the major financial markets. Again,
a fraction of the companies exhibit a behavior similar to that
of the major financial markets. A clear-cut distinction is always difficult,
but we find three groups according to the outlier criterion:
\begin{itemize}
\item {\bf Obvious outliers:} General Electric, Intel, Exxon-Mobil, Oracle,
Wall-Mart, IBM, HP, Sun Microsystems, SBC, Coca-Cola, Procter \& Gamble and 
Medtronic.

\item {\bf Difficult to classify in the first group :} AT\&T, Citigroup, 
Applied Materials and General Motors.

\item {\bf Clearly no outliers:} Microsoft, Cisco, Lucent, Texas Instrument,
Merck, EMC, Pifzer, AOL, MCI WorldComm., Qualkomm, JDS Uniphase, 
American Home Products and Ford.
\end{itemize}

This means that $12$ companies belong to the first catagory, $13$ to
the last and only $4$ companies on the borderline. 
Hence, it is difficult to conclude with the same force as for the drawdowns
about the existence of outliers. The concept of large drawups as outliers
seems however a persuasive feature of about half the markets. 
Another interesting observation is that the fits with 
the stretched exponential are consistent with a value for the exponent 
$z \approx 1$ (albeit with significant fluctuations
from market to market). The two observations
may be related. Again, it is interesting to observe that the two companies,
America Online and JDS Uniphase, which have by far the highest rate of drawdowns
above $15\%$ also have by far the highest rate of drawups above $15\%$, 
as shown in table \ref{dutable}. In fact, this ranking holds approximately
for most of the companies. Figure \ref{Du-DDregress} shows the number of
drawups of amplitude larger than $15\%$
as a function of the number of drawdowns larger than $15\%$
for all the companies analyzed
here and shown respectively in tables \ref{dutable} and \ref{ddtable}.
The linear regression is good with a $R$-statistics of $0.97$ and a correlation coefficient
of $0.62$. In the linear regression, we have allowed the value $M1$ at the origin to be non-zero.
However, the fit finds $M1$ to have a negligible value: vanishing draw downs come with
vanishing draw ups. 

Quantitatively, there
are on average $0.94$ drawdowns per year per company
of amplitude equal to or larger than $15\%$, compared to $1.92$ drawups 
per year per company of amplitude equal to or larger than $15\%$. 
Large drawups are twice as frequent and thus to a much lesser extent
to be characterised as outliers.
The difference cannot be attributed to a difference in the 
size of the two sets. It may be related to the very different value
given to them by the market:
drawups are cheered upon by the market and the media, whereas \dds
are not considered well\footnote{Unless you are ``short'' due to some insight
not shared by the vast majority of the market players.}. Hence, both from the
point of view of risk management and of market psychology, there is 
a strong asymmetry between drawups and drawdowns.

\section{Synthetic tests and statistical significance}

To further establish the statistical confidence with which we can conclude
that the largest drawdowns are outliers, for each financial time series,
we have reshuffled the daily returns 1000 times 
and hence generated 1000 synthetic data sets for each. This
procedure means that 1000 synthetic data will have exactly the same
distribution of daily returns as each time series. However, higher order correlations apparently
present in the largest drawdowns are destroyed by the reshuffling. This
surrogate data analysis of the distribution of drawdowns has the advantage
of being {\it non-parametric}, {\it i.e.}, independent of the quality of fits
with a model such as the stretched exponential. 

In order to compare the
distribution of drawdowns obtained for the real data and for the synthetic data 
for each of the indices, of the currencies and of companies
named in the first column of tables \ref{surrddmarkets} and \ref{surrddcomp}, 
a threshold given by the second column
has been chosen by identifying the point of breakdown from the stretched exponential, {\it i.e.}, the crossover point from bulk to outliers:
the first point after this crossover point
is then the smallest outlier and the threshold is the integer value
of that drawdown. The
third column gives the number of drawdowns above the threshold in the 
true data. The fourth column gives the number of
surrogate data sets with $0,1, 2, 3, ...$ 
drawdowns larger than the threshold. The last column quantifies
the corresponding confidence level.

The results for the
financial market indices, for the currencies and for gold
exhibit a very high statistical confidence level about $99\%$, as shown 
in table \ref{surrddmarkets}.
The situation is more dispersed for companies as shown 
in table \ref{surrddcomp}: out
of the $25$ companies presented in the table, six (resp. fourteen) companies 
gives a confidence level higher than $98\%$ (resp. $80\%$).
We expect and preliminary results indicate that generalizing drawdowns
to take into account short upward moves in an otherwise downward trend
will enhance this statistical significance for the companies. This
will be reported in a future work.

The exponential null-hypothesis which seems to be 
valid for companies provides another measure of the departure
of the outliers from the bulk of the distributions. As shown in 
Appendix B, the typical fluctuations of the largest drawdowns
are expected to be given by the expression
(\ref{mjalanzka}): the expected fluctuations of the 
largest drawdowns observed in a finite sample of $N$ events should thus be of the
order of the typical drawdown size $\chi$. Since $\chi$ is typically found of the
order of a few percent at most and no more than about $6\%$ for the ``wilder''
companies, the observed large values of the outliers above
the extrapolation of the exponential fit by $10\%$ up to 
about $20\%$ is many times  
larger than the variations predicted by the exponential null-hypothesis.

\section{Summary and discussion}

This paper has investigated the distribution of drawdowns
defined as losses from the last local maximum to the next local minimum
and of drawups defined as gains from the last local minimum to the next
maximum. Drawdowns are particularly interesting because they provide
a more natural measure of real market risks than the variance or other
centered moments of daily (or other fixed time scale) distributions of returns. 
The analyzed time series consist in most of the major world financial
indices, the major currencies, gold and the twenty 
largest U.S. companies in terms of capitalisation as well as nine
others chosen randomly.
We have found the following facts.
\begin{enumerate}
\item Large drawups of more than $15\%$ occur approximately twice as
often as large drawdowns of similar amplitudes.

\item The bulk ($98\%$) of the drawdowns and drawups are very well-fitted
by the exponential model, which is the natural null-hypothesis for the class
of exponential or superexponential return distributions, when assuming 
independent increments. We have also used a generalization called the stretched
or Weibull exponential model $\propto \exp(-|x/\chi|^z)$ which allowed
us to account for the subexponential class of return distributions (see
Appendix A).

\item The typical scale $\chi$ (proportional to the standard
deviations of the price variations) and the exponent $z$ provide
two useful measures of the size of drawdowns and their rate of occurrence:
the larger $\chi$ is, the larger are the drawdowns and drawups. The smaller $z$ is,
the fatter is the tail of the distribution controlling large fluctuations.

\item As expected, currencies have the smallest typical fluctuations
both for drawdowns and drawups measured by $\chi=0.79-1.29\%$ but have relatively 
fat tails $z =0.84-0.91$. For indices, fluctuations are larger 
with $\chi=1.05-2.1\%$
with $z$ for drawdowns in the same range as the currencies (except CAC40 which is
compatible with a pure exponential $z=1$). In constrast, drawups have the same range
of $\chi$ but their exponents are larger and are compatible with $z=1$.
Companies have much larger drawdowns in general with $\chi=1.91-7.61\%$.
The exponents $z$ are compatible with the null-hypothesis $z=1$ both 
for drawdowns and drawups.

\item Since the major financial indices (except the CAC40), gold and the 
exchange rates are characterized by an exponent $z<1$, this is compatible with 
a daily return distribution in the subexponential class, as discussed in Appendix A,
which is characterized by fat tails. In contrast, the distribution of drawdowns
of the large US companies are compatible with $z \approx 1$, compatible with 
a distribution of daily returns in the exponential or superexponential class.

\item The remaining $1-2\%$ of the largest drawdowns are not at all explained
by the exponential null-hypothesis or its extension 
in terms of the stretched exponential.
Large drawdowns up to three times larger than expected from the null-hypothesis
are found to be ubiquitous occurrences of essentially all the times series
that we have investigated, the only noticeable exception being the French index
CAC40. We call these anomalous drawdowns ``outliers''. This emphasizes that
large stock market drops (including crashes) cannot be accounted for 
by the distribution of returns characterising the smaller market moves. They thus
belong to a different class of their own and call for a specific amplification
mechanism. 

\item About half of the time series show outliers for the drawups. The drawups 
are thus different statistically from the drawdowns and constitute a less
conspicuous structure of financial markets.
\end{enumerate}

The most important result is the demonstration that the very largest drawdowns
are outliers. This is true notwithstanding the fact that the very largest
daily drops are {\it not} outliers, most of the time. Therefore, the anomalously
large amplitude of the drawdowns can only be explained
by invoking the emergence of rare but sudden persistences of 
successive daily drops, with in addition correlated amplification of the
drops. Why such successions of correlated daily moves occur
is a very important question with 
consequences for portfolio management and systemic risk, to cite
only two applications. In previous works, we have argued that the very largest
drawdowns, the financial crashes, are the collapse of speculative bubbles
and result from specific imitation and speculative behavior. A model with 
a mixture of rational agents and noisy imitative traders accounts for the stylized
observations associated with crashes (Johansen and Sornette, 1999a,b; 2000a;
Sornette and Andersen, 2001). In 
this model, the outlier crashes are indeed understood as special events corresponding
to the sudden burst of a bubble. 

Our main result that there is some (transient) correlation across
daily returns adds to the existing empirical literature on market
microstructure effects.  For instance, it is known that emerging stock markets
have large positive autocorrelation.  In this case, this is not an inefficiency but rather
reflects the fact that stocks are infrequently traded, which induces first-order
autocorrelations.  Similarly, during the crash of Oct. 1987, there were order
imbalances that carried from one day to another and certainly exacerbated the
crash. It would be interesting to investigate whether arbitrage opportunities
can be obtained in the ``outlier'' regime. In any, they must be rather small
due to the infrequent occurrences of these anomalous events.

The implications of our results for risk management are the following. While
the literature of the Value-at-Risk and on extreme value theory (EVT)
focus its analysis on one-day extreme events occurring once a year, once
a decade or once a century, we have shown that this may not be the most
important and most relevant measure of large risks: large losses occur
often as the result of transient correlations leading to runs of 
cumulative losses, the drawdowns. The existence of these transient
correlations make large drawdowns much more frequent than expected from the
estimation of the tail of return distributions, assuming independence between
successive returns. In addition, the fact that the very large drawdowns are
``outliers'' show that the characterization of the tail of their distribution
cannot be based on standard techniques extrapolating from smaller values.
In sum, our results suggest to reconsider the large risks as fundamentally
composite events, which require new statistical tools based on a multivariate
description of returns at different times and variable time scales.

In the spirit of Bacon in Novum Organum about 
400 years ago, ``Errors of Nature, Sports and Monsters correct the 
understanding in regard to ordinary things, and reveal general forms. For 
whoever knows the ways of Nature will more easily notice her deviations; 
and, on the other hand, whoever knows her deviations will more accurately 
describe her ways,'' we propose that drawdown outliers reveal 
fundamental properties of the stock market.

In future works, it will be interesting to examine the variations of the statistics of
drawdowns, of drawups, of the typical scale $\chi$ and of the exponent $z$
in different parts of the industry to investigate whether 
industry specificities may lead to characteristic structures for the drawdowns, for the
existence, size and rate of outliers.

\vskip 0.5cm
We acknowledge encouragements from D. Stauffer and thank the referee and P. Jorion as the 
editor for constructive remarks. DS acknowledges stimulating 
discussions with T. Lux and V. Pisarenko.

\newpage

\section{Appendix A: The distribution of drawdowns for independent price variations}

In this appendix, we show that the distribution of drawdowns for independent
price increments $x$ is asymptotically an exponential when the distribution
of $x$ does not decay more slowly than an exponential, {\it i.e.}, belong to the class
of exponential or super-exponential distributions. In contrast, for sub-exponentials
(such as stable L\'evy laws, power laws and stretched exponentials), the 
tail of the distribution of drawdowns is asymptotically the same as the 
distribution of the individual price variations.

\subsection{The equation giving the distribution of drawdowns}

Consider the simplest and most straightforward definition of
a drawdown, namely the loss in percent from a maximum to the following 
minimum. For instance, if the signs of returns over 12 consecutive days 
are $++--+-+----+$, the first drawdown lasts two days, the second
drawdown lasts one day and the third one lasts four days. With this
definition and under the null hypothesis of uncorrelated consecutive 
price returns, the probability density function $P(D)$ to observe a drawdown of a given 
magnitude $D$ is
\be
P(D) = {p_+ \over p_-} \sum_{n=1}^{\infty} \int_{-\infty}^0 dx_1 ~ p(x_1)
... \int_{-\infty}^0 dx_n ~ p(x_n)
\delta \left(D-\sum_{j=1}^n x_j\right)~, \label{ooiukioa}
\ee
where
\be
p_+ =1-p_-=\equiv \int_0^{+\infty} dx ~ p(x)
\ee
is the probability to observe a positive price variation and the term $p_+/p_-$
ensure the normalization of
$P(D)$. The Dirac function $\delta \left(D-\sum_{j=1}^n x_j\right)$ ensures that the
sum over $n$ in the r.h.s. of (\ref{ooiukioa})
is over all possible run lengths $n$ of consecutive losses $x_1, ..., x_n$ that sum up to a
given $D$.

\subsection{An exactly solvable case: the family of Gamma distributions}

Expression (\ref{ooiukioa}) can be solved explicitely for the family of
Gamma density distributions defined here for $x<0$ as
\be
g_q(x) = p_- ~\alpha {(\alpha |x|)^{q-1} \over (q-1)!}~ e^{-\alpha |x|}~,
\ee
where $q=1, 2, 3,...$ can take any positive integer value. The prefactor $p_-$ ensures
that $\int_{-\infty}^0 g_q(x) = p_-$.
The solution relies on the fact that the distributions $g_q(x)$ are stable
in family with respect to convolution. Thus, for a given $q$, 
expression (\ref{ooiukioa}) gives
\be
P(D) = {p_+ \over p_-} ~\alpha~\sum_{n=1}^{\infty}  p_-^n {(\alpha D)^{nq-1} 
\over (nq-1)!}~ e^{-\alpha D} = {p_+ \over p_-} 
~\alpha~ e^{-\alpha D}~p_-^{1 \over q}~\sum_{n=1}^{\infty} 
{\left(p_-^{1 \over q}~ \alpha D \right)^{nq-1}  \over (nq-1)!}~
\label{hganlam}
\ee
In deriving (\ref{hganlam}), we have used the fact that $\int_{-\infty}^0 dx_1 ~ p(x_1)
... \int_{-\infty}^0 dx_n ~ p(x_n) \delta \left(D-\sum_{j=1}^n x_j\right)$
is nothing but $g_q(x)$ convoluted $n$ times with itself.

\begin{enumerate}
\item For $q=1$, {\it i.e.}, $g_1(x) = p_- ~\alpha~ e^{-\alpha |x|}$, we find
\be
P(D) = {p_+ \over p_-}~\alpha~e^{-|D|/D_1}~,
\ee
where 
\be
1/D_1 = \alpha (1-p_-)~.
\ee

\item  For $q=2$, {\it i.e.}, $g_2(x) = p_- ~\alpha (\alpha |x|) ~ e^{-\alpha |x|}$, we find
\be
P(D) = {p_+ \over p_-}~\alpha~\sqrt{p_-}~e^{-\alpha |D|}~\sinh (\sqrt{p_-}~\alpha |D|)
~~~\rightarrow ~~~ {p_+ \over 2 p_-}~\alpha~\sqrt{p_-}~~e^{-|D|/D_2}~,~~~~{\rm for}~|D|>D_2~,
\ee
where 
\be
1/D_2 = \alpha (1-\sqrt{p_-})~.
\ee

\item  For $q=3$, {\it i.e.}, $g_3(x) = p_- ~\alpha {(\alpha |x|)^2 \over 2} ~ e^{-\alpha |x|}$, we find
\be
P(D) = {p_+ \over p_-}~\alpha~p_-^{1 \over 3}~e^{-\alpha |D|}~{1 \over 3}
\left[ \exp \left( X \right) + j \exp \left( j X \right) + j^2 \exp \left( j^2 X \right)\right]~,
\ee
where $X \equiv  p_-^{1 \over 3}~ \alpha |D|$ and $j=e^{i 2\pi/3}
= -{1 \over 2} + i {\sqrt{3} \over 2}$ is the first cubic root of $1$. This leads to
\be
P(D) \rightarrow {p_+ \over 3p_-}~\alpha~p_-^{1 \over 3}~e^{-|D|/D_3}~,~~~~{\rm for}~|D|>D_3~,
\ee
where 
\be
1/D_3 = \alpha (1-p_-^{1 \over 3})~.
\ee

\item For a general $q>1$, we find
\be
P(D) = {p_+ \over p_-}~\alpha~p_-^{1 \over q}~e^{-\alpha |D|}~{1 \over q}
\left[ \exp \left( X \right) + \omega \exp \left( \omega X \right) + \omega^2 
\exp \left( \omega^2 X \right) +...+ \omega^{q-1} \exp \left( \omega^{q-1} X \right) \right]~,
\ee
where $X \equiv  p_-^{1 \over q}~ \alpha |D|$ and $\omega=e^{i 2\pi/q}$
is the first $q$-th root of $1$,{\it i.e.}, $\omega^q=1$. This leads to
\be
P(D) \rightarrow {p_+ \over qp_-}~\alpha~p_-^{1 \over q}~e^{-|D|/D_q}~,~~~~{\rm for}~|D|>>D_q~,
\label{ngnlflala}
\ee
where 
\be
1/D_q = \alpha (1-p_-^{1 \over q})~.   \label{ghwlw}
\ee
Note that, for large $q$, $D_q  \to q/\alpha \ln(1/p_-)$, hence the tail of the
distribution of drawdowns becomes fatter as $q$ increases. This reflects simply 
the fact that $g_q(x)$ develops a fatter tail as $q$ increases. In particular,
the expectation of $x$ conditioned to be negative increases with $q$ as
\be
\langle x\rangle_- = - {q \over \alpha}~.   \label{ghngal}
\ee

\end{enumerate}

These results show that the tail of the distribution of drawdowns for the Gamma family
is an exponential. In the sequel, this result is shown to hold asymptotically 
for general distributions $p(x)$  when their tails
decay no slower than an exponential (up to 
algebraic factors).

\subsection{General asymptotic exponential law for exponential or super-exponential
distributions}

Taking the Laplace transform of (\ref{ooiukioa}) gives the expression of
the characteristic function
${\hat P}(k)$ obtained after the summation of the infinite series:
\be
{\hat P}(k) = {p_+ \over p_-} ~{{\cal P}(k) \over 1- {\cal P}(k)}~,
\label{ooiuafdkioa}
\ee
where
\be
{\cal P}(k) \equiv \int_{-\infty}^0 dx ~ p(x) ~ e^{kx} ~.
\label{uajhaka}
\ee
Notice that ${\cal P}(k)$ is a modified characteristic function of $p(x)$
truncated to negative price
variations.
Expression (\ref{ooiuafdkioa}) can also be rewritten as
\be
{\hat P}(k) = {1 \over 1 - {1\over p_+} ~{{\cal P}(k) - {\cal P}(0) \over
{\cal P}(k) }}~,  \label{nhaka}
\ee
where we have used the definition
${\cal P}(0) \equiv p_-$. 

If the distribution $P(x)$ does not decay more slowly than an exponential
(the following results thus exclude the class of sub-exponential
distributions, such as power laws, stable L\'evy laws and stretched
exponentials discussed below),
we can expand $\left({\cal P}(k) - {\cal P}(0)
\right) /{\cal P}(k) $
for small $k$ (corresponding to large $|D|$'s as
\be
{{\cal P}(k) - {\cal P}(0) \over {\cal P}(k) } = k {d{\ln \cal P}(k) \over
dk}|_{k=0} + {\cal O}(k^2) ~,  \label{gnlnvqlvf}
\ee
which together with (\ref{nhaka}) yields
\be
{\hat P}(k) = {1 \over 1 + kD_0}~,    \label{jajjan}
\ee
where
\be
D_0 = - {\langle x\rangle_- \over p_+}~,  \label{nnvd}
\ee
and
\be
\langle x\rangle_- = \int_{-\infty}^{0} dx ~x~ p(x)  \label{hfbkba}
\ee
is the average price variation conditioned to be negative: in other word,
it is the average size of the negative variations. Estimating the
two parameters $\langle x\rangle_-$ and $p_+$ from the empirical data
for the DJIA, the Nasdaq composite index and the S\&P500 index gives
respectively  $D_0 = 0.0134, 0.01265$ and $0.0115$.

The distribution whose characteristic function is (\ref{jajjan}) is nothing
but the exponential
function
\be
P(D) = {e^{-|D| \over D_0} \over D_0}~,  \label{jhdbjas}
\ee
where $D_0>0$ is the typical amplitude of
the drawndown amplitude distribution. This derivation shows that the 
tail of the drawndown distribution defined as runs of negative
returns is generically an exponential function. For approximately
symmetric distributions of daily returns, $p_+=p_-=1/2$. Formula (\ref{nnvd})
shows that the typical size $D_0$ of drawdowns is $2$ times the average daily drop.

The general approximate result (\ref{jhdbjas}) retrieves the asymptotic regime 
(\ref{ngnlflala}) with (\ref{ghwlw}) for the Gamma family as follows. Using (\ref{ghngal})
in (\ref{nnvd}) gives $D_0 = q/\alpha p_+$ that should be compared with the
exact result (\ref{ghwlw}). In the limit of large $q$, 
$D_q  \to q/\alpha \ln(1/p_-)$. And for not too large $p_+$, $\ln(1/p_-) = \ln(1/(1-p_+))
\approx p_+$. In this limiting case, the
general approximate result (\ref{jhdbjas}) thus retrieves the exact asymptotic regime 
(\ref{ngnlflala}) with (\ref{ghwlw}). However, this comparison warns us that
the expansion of $\left({\cal P}(k) - {\cal P}(0) \right) /{\cal P}(k) $
for small $k$ leading to (\ref{jajjan}) provides only an approximation.

We retrieve the same exponential distribution function
semi-quantitatively by considering that negative price variations are of a 
fixed size $\langle x\rangle_-$. Then, the probability to observe a 
drawdown of amplitude $D=n \langle x\rangle_-$ is simply $p_+ p_-^n$, 
which gives (\ref{jhdbjas}) with a slightly modified estimation of the 
typical drawdown $D_0 = - {\langle x\rangle_- \over \ln 1/p_-}$.
This exponential distribution (\ref{jhdbjas}) of ``current drawdowns''
should not be mistaken with the distribution of the maximum drawdowns 
defined as the largest depression from some arbitrary past price, which 
can also be shown to have an exponential tail with a typical value $D_0$ 
determined from the equation
(Bouchaud and Potters, 1997; see Feller, 1971, p.402; Sornette and Cont, 1997)
\be
\int_{-\infty}^{+\infty} dx~ p(x)~ e^{-{D \over D_0}} = 1~.
\ee
The exponential distribution (\ref{jhdbjas}) thus constitutes our null
hypothesis for exponential or superexponential distributions of returns.

With respect to durations of drawdowns, we
expect them to be exponentially distributed
from the assumed independence of signs of successive price variations.
The probability that a drawdown lasts $n$ consecutive days, {\it i.e.}, is a 
run of length $n$, is
$p_+ p_-^n \propto \exp\left(n \ln p_-\right)$. The typical duration of 
a drawdown is thus
\be
n_0 = {1 \over \ln {1\over p_-}}~.  \label{jfjslls}
\ee
Since most markets exhibit very weak asymmetries, the probability $p_-$
for a negative price variation is close to $1/2$ leading to $n_0 = 1/\ln 2
\approx 1.44$ days. The exponential distribution 
of the duration of drawdowns and the typical value (\ref{jfjslls})
turns out to be in good agreement with our data analysis. 
A similar result has
been first derived by von Mises (1921) who showed that the number of long
runs of given length was approximately distributed according to the Poisson law
for large samples. Mood (1940) has shown that the body of the distribution of run lengths
is asymptotically Gaussian.

\subsection{Next order term of the distribution of drawdowns}

To improve on the asymptotic result (\ref{jhdbjas}) obtained by
keeping only the linear term in $k$ in the denominator of (\ref{nhaka}),
we now keep all terms up to second order in $k$:
\be
{{\cal P}(k) - {\cal P}(0) \over {\cal P}(k)} = {k {\cal P}'(0) +{1 \over 2} 
k^2 {\cal P}''(0)  \over {\cal P}(0) + k {\cal P}'(0) +{1 \over 2} 
k^2 {\cal P}''(0)} = {k {\cal P}'(0) \over {\cal P}(0)}~
{1 + {k\over 2} {{\cal P}''(0) \over {\cal P}'(0)} \over
1 + k {{\cal P}'(0) \over {\cal P}(0)}} = 
{k {\cal P}'(0) \over {\cal P}(0)}
\left(1 + k \left[ {1\over 2} {{\cal P}''(0) \over {\cal P}'(0)} -
{{\cal P}'(0) \over {\cal P}(0)}\right]\right)~.  \label{gtqktg}
\ee
This expression (\ref{gtqktg}) refines (\ref{gnlnvqlvf}).
By replacing in (\ref{nhaka}), we thus obtain
\be
{\hat P}(k) = {1 \over 1 + kD_0(1+kD_1)}~,    
\label{jajjaqqrn}
\ee
where $D_0$ is still given by (\ref{nnvd}). Noting that
\be
{\cal P}''(0) = - \langle x^2 \rangle_- = - \int_{-\infty}^{0} dx ~x^2~ p(x) ~,
\label{ghhgqlvm}
\ee
we obtain $D_1$ from (\ref{gtqktg}) as
\be
D_1 = -  \left({\langle x^2\rangle_- \over 2\langle x\rangle_-}
+ {\langle x \rangle_- \over p_-}\right)~,  \label{nngfqgvd}
\ee
which is positive because $\langle x^2\rangle_->0$ and $\langle x\rangle_-<0$.

We note that $D_1>D_0/4$ in general since $p_+ \approx p_- \approx 1/2$ which
implies $D_1 > {-\langle x \rangle_- \over p_-} > {-\langle x \rangle_- \over 4p_+} = D_0/4$.

The distribution whose characteristic function is (\ref{jajjaqqrn}) is then an
oscillatory exponential function
\be
P(D) = e^{-2|D|\over D_0} ~[A \cos(\omega |D|) + B \sin(\omega |D|)] \label{jhdbfhgjas}
\ee
where
\be
\omega = {1 \over 2} \sqrt{4 D_0 D_1-D_0^2}~.
\ee
Note that, for $D_1=0$, $\omega$ become imaginary and the oscillatory term
becomes a pure exponential which, together with the first term $e^{-2|D|\over D_0}$
retrieves (\ref{jhdbjas}) as it should. 

The result (\ref{jhdbfhgjas}) deriving from 
the fact that $D_1>D_0/4$ shows that the correction to the exponential (\ref{jhdbjas})
tends to give an upward curvature, suggesting a slower decay than
predicted by keeping only the first term in 
powers of $k$ in the expansion (\ref{gnlnvqlvf}).
However, the solution
(\ref{jhdbfhgjas}) does not hold for large $D$ (it would predict a
probability becoming negative) and it is bound to fail
due to the effect of the neglected terms.

More generally, we can express 
${\cal P}(k)$ as a sum over moments (which exist and are equivalent to the
knowledge of the distribution of returns $P(x)$ if it decays 
no slower than an exponential)
\be
{\cal P}(k) = p_- + \sum_{n=1}^\infty {k^n m_n \over n!}
\ee
where 
\be
m_n = \int_{-\infty}^0 dx ~p(x)~x^n ~.
\ee
Then, expression (\ref{nhaka}) gives
\be
{\hat P}(k) = {1 \over 1 - {1\over p_+} {\sum_{n=1}^\infty {k^n m_n \over n!}
\over p_- + \sum_{n=1}^\infty {k^n m_n \over n!}}}~,  \label{naahaka}
\ee
Obtaining the asymptotic tail of ${\hat P}(D)$ requires a resummation of this
expression.

\subsection{Another exactly soluble case: stable laws and power laws}

Consider the family of stable asymmetric L\'evy laws restricted here to $x<0$
\be
p(x) = p_- ~N~L_{\beta=-1, \gamma, \alpha}(x)
\ee
where the asymmetric $L_{\beta=-1, \gamma, \alpha}(x)$ has the characteristic function
\be
{\hat L}_{\beta=-1, \gamma, \alpha}(k) = \exp \left[ i \gamma k - C_{\alpha,\beta=-1} |k|^{\alpha} 
 \right]~.
\ee
With $N=\int_{-\infty}^0 L_{\beta=-1, \gamma, \alpha}(x)$, 
the prefactor $p_-$ is the total probability
for a price variation to fall in the interval $-\infty$ to $0$. Note that 
$N=1$ for $0<\alpha <1$ (the asymmetric L\'evy law 
is zero for positive arguments) while $N<1$ for $1 \leq \alpha <2$ (the L\'evy law is non-zero
for positive arguments) [Sornette, 2000].

The L\'evy distribution $L_{\beta, \gamma, \alpha}(x)$  is stable under $n$ convolutions: 
the distribution of the sum $S_n$ of $n$ independent variables distributed each according
to $L_{\beta, \gamma, \alpha}(x)$  is exactly $L_{\beta, \gamma, \alpha}\left(
{S_n - \gamma n \over n^{1/\alpha}}\right)$. This allows us to transform
expression (\ref{ooiukioa}) as
\be
P(D) \approx {p_+ \over p_-} \sum_{n=1}^{\infty}  p_-^n ~L_{\beta=-1, \gamma, \alpha}\left(
{|D| - \gamma n \over n^{1/\alpha}}\right)~.
\label{ahganlaaam}
\ee
This expression (\ref{ahganlaaam}) is exact for $0<\alpha <1$ for all $|D|$'s, while
it is only valid for large $|D|$'s for $1 \leq \alpha$.
This allows us to obtain the tail behavior of $P(D)$ by summing over the tails of
each term in the sum. This is warranted as the contributions of the bulk (non-power
law tail) of the L\'evy laws gives exponentially small corrections for asymptotically
large $|D|$'s. Using $L_{\beta=-1, \gamma, \alpha}(|D|) \sim C_{\alpha} /|D|^{1+\alpha}$
for $|D|>> C_{\alpha}^{1/\alpha}$ where $C_{\alpha}$ is a scale factor, we obtain
\be
P(D) \to {p_+ \over p_-} {C_{\alpha} \over |D|^{1+\alpha}}
\sum_{n=1}^{\infty}  n p_-^n = 
{C_{\alpha} \over p_+}~ {1 \over |D|^{1+\alpha}}~, ~~~~
{\rm for}~|D|>> C_{\alpha}^{1/\alpha}~.    \label{gnnqlqlq}
\ee

This result (\ref{gnnqlqlq}) holds more
generally asymptically for large $|D|$'s for distributions $p(x)$ which have power law tails
with exponents $\alpha >2$ outside the stable law regime.

\subsection{The general class of sub-exponentials}

Let us denote ${\bar P}_x(x)$ the tail distribution for negative returns $x$. 
$P_x$ is a sub-exponential if (see [Sornette, 2000] and references therein)
\be
{\rm lim}_{|x| \to \infty}~~{{\bar P}_x^{n*}(x) \over {\bar P}_x(x)} = n
~~~~~~{\rm for~all}~n \geq 2~,
\ee
where ${\bar P}_x^{n*}(x)$ is the tail distribution of the $n$-fold convolution of
$p_x(x)$. In words, the sum of $n$ subexponential random variables is large if and only
if their maximum is large. In other words, the sum of $n$ subexponential random
variables behaves asymptotically for large values as the largest of them.
Rewriting (\ref{ooiukioa}) as
\be
{\bar P}(D) = {p_+ \over p_-} \sum_{n=1}^{\infty} {\bar P}_x^{n*}(D)
\ee
and replacing ${\bar P}_x^{n*}(D)$ by its asymptotic value ${\bar P}_x^{n*}(D) 
\to n {\bar P}_x(D)$ for large $|D|$, leads to
\be
{\bar P}(D) \sim {1 \over p_+}~ {\bar P}_x(D) ~.   \label{bhgnw}
\ee
Expression (\ref{bhgnw}) thus shows that the tail of the distribution of 
drawdowns for large $|D|$ has the same functional shape as the tail of the
subexponential
distribution of price variations. This general result retrieves the
result (\ref{gnnqlqlq}) obtained for stable L\'evy laws and for power laws
which are indeed subexponential distributions. 

But it extends it to other
functional shapes. 
For instance, Laherr\`ere and Sornette [1998] have proposed to quantify the
distribution of price variations and of price returns by the family of 
stretched exponentials (\ref{stretched}), which has also been used
in an extension of portfolio theory [Sornette et al., 2000; Andersen and Sornette, 2001].
This family (\ref{stretched}) with $z<1$ belongs to the class of subexponentials
for which the result (\ref{bhgnw}) holds. Therefore, if the distribution of price
variations is a stretched exponential, so will be the tail of the distribution of
drawdowns.

\newpage

\section{Appendix B:  Confidence interval for 
large drawdowns distributed with for an exponential law}

Let us consider the exponential distribution $P(D)$ given by
(\ref{jhdbjas}) and let us assume
that we have observed $N$ drawdowns. We would like to determine the
typical amplitude of
the fluctuations of these drawdowns, and especially the fluctuations of
the largest ones. Here, we consider the absolute values or amplitudes
of drawdowns and thus work with positive numbers.
We rank them by descending values $D_1 > D_2 > ... > D_N$ and
express the
probability $P(D_n)$ that the $n$-th rank $D_n$ takes a given value in the
interval of width $dD_n$:
\be
P(D_n) dD_n =  {N\choose n} \biggl(1 - \int_{D_n}^{\infty} P(D) dD
\biggl)^{N-n}
{n\choose 1} \biggl(\int_{D_n}^{\infty} P(D) dD \biggl)^{n-1} P(D_n) dD_n~.
\ee
The most probable $n$-th rank $D_n^*$ maximizes $P(D_n)$ and is given by
\be
D_n^* = D_0 \ln{N \over n}~.   \label{jfbjabaq}
\ee
The half-width (defined as the deviation from $D_n^*$ that halves $P(D_n)$)
is given by
\be
\Delta D_n \approx {D_0 \over \sqrt{n}}~,   \label{mjalanzka}
\ee
for small ranks $n << N$, {\it i.e.}, for the largest drawdowns.

This result (\ref{mjalanzka}) shows that the expected fluctuations of the 
largest drawdowns observed in a finite sample of $N$ events is of the
order of the typical drawdown size $D_0$. Since $D_0$ is typically found of the
order of a few percent at most and no more than about $6\%$ for the ``wilder''
companies, the observed large values of the outliers above
the extrapolation of the exponential fit by $10\%$ up to 
about $20\%$ is many times  larger than the variations predicted by the null-hypothesis.

\newpage

{\bf References}

\vskip 0.5cm

Andersen, J.V. and D. Sornette, 2001,
Have your cake and eat it too: increasing returns while lowering large risks!
Journal of Risk Finance 2 (3), 70-82.

Arneodo, A., Muzy, J.F. and Sornette, D., 1998,
``Direct'' causal cascade in the stock market, European Physical Journal B
2, 277-282.

Ballocchi, G., M. M. Dacorogna, R. Gencay, 1999,
Intraday Statistical Properties of Eurofutures by Barbara Piccinato,
Derivatives Quarterly 6, 28-44.

Barber, B.M. and Lyon, J.D., 1997,
Detecting long-run abnormal stock returns: The empirical power and
specification of test statistics, Journal of Financial Economics 43, N3,
341-372.

Bassi, F., P.Embrechts, and M.Kafetzaki (1998) 
Risk Management and Quantile Estimation, in: Adler, R.J., R.E.Feldman, 
M.Taqqu, eds., A Practical Guide to Heavy Tails, Birkhauser, Boston, 111-30.

Bouchaud, J.-P. and Potters, M., 1997,
Th\'eorie des Risques Financiers, Al\'ea Saclay, Paris, Diffusion Eyrolles.

Breymann, W., S. Ghashghaie and P. Talkner, 2000, A stochastic cascade model for
FX dynamics, preprint cond-mat/0004179 

Campbell, J.Y., A.W. Lo, A.C. MacKinlay, 1997,
The econometrics of financial markets, Princeton, N.J. : Princeton
University Press.

Clark, P.K., 1973, A subordinate stochastic process model with finite
variance for speculative prices, Econometrica 41, 135-155.

Dacorogna, M.M., Gauvreau, C.L., M\"uller, U.A., Olsen, R.B. et al., 1996,
Changing time scale for short-term forecasting in financial markets,
Journal of Forecasting 15, 203-227.

Dacorogna, M.M., U.A. M\"uller, R.B. Olsen, O.V. Pictet, 1998,
Modelling Short-Term Volatility with GARCH and HARCH Models,
in ``Nonlinear Modelling of High Frequency Financial Time Series,'' by C.
Dunis, B. Zhou (John Wiley \& Sons).

Diebold, F.X., Schuermann, T. and Stroughair, J.D. (2001)
Pitfalls and opportunities in the use of extreme value theory in risk
management, preprint.

Embrechts, P., C.P.Kluppelberg, and T.Mikosh (1997) Modelling Extremal Events, 
Springer-Verlag, Berlin, 645 pp.

Feller, W., 1971, An introduction to probability theory and its
applications, Vol. II, second edition
John Wiley and sons, New York.

Geman, H. and An\'e, T., 1996, Stochastic subordination, RISK, september.

Ghashghaie, S., W. Breymann, J. Peinke, P. Talkner and Y. Dodge, 1996,
Turbulent cascades in foreign exchange markets, Nature 381, 767-770.

Gopikrishnan, P., Meyer, M., Amaral, L.A.N. \& Stanley, H.E., 1998,
Inverse Cubic Law for the Distribution of  Stock Price Variations,
European Physical Journal B 3, 139-140.

Grossman, S, and Z. Zhou, 1993, Optimal investment strategies for 
controlling drawdowns, Mathematical Finance 3 (3), 241-276.

Guillaume, D.M., Pictet, O.V., M\"uller, U.A. and Dacorogna, M.M., 1995,
Unveiling nonlinearities through time scale transformations, preprint
O\&A Research Group, $http://www.olsen.ch/library/research/oa_working.html$

Guillaume, D.M., Dacorogna, M.M., Dav\'e, R.R,, M\"uller, J.A., Olsen, R.B. 
\& Pictet, O.V., 1997, From the Bird's Eye to the Microscope: 
A Survey of New Stylized Facts of the intra-daily Foreign Exchange Markets, 
Finance and Stochastics 1, 95-129.

Hsieh, D.A., 1989, Testing for nonlinear dependence in daily foreign 
exchange rates, Journal of Business 62, 339-368.

Johansen, A. and D. Sornette, 1998, Stock market crashes are outliers, 
European Physical Journal B 1, 141-143.

Johansen, A. and D. Sornette, 1999a,
Critical Crashes, RISK 12 (1), 91-94.

Johansen, A., D. Sornette and O. Ledoit, 1999b,
Predicting Financial Crashes using discrete scale invariance,
Journal of Risk 1 (4), 5-32.

Johansen, A., O. Ledoit and D. Sornette, 2000a,
Crashes as critical points,  International Journal of Theoretical and Applied Finance
3 (2),  219-255.

Johansen, A. and D. Sornette, 2000b, 
The Nasdaq crash of April 2000: Yet another example of
log-periodicity in a speculative bubble ending in a crash,
European Physical Journal B 17, 319-328. 

Knuth, D.E., 1969, The art of computer programming, vol.2, 1-160,
Addison-Wesley Publ.

Laherr\`ere, J. and D. Sornette, 1998, Stretched exponential distributions
in Nature and Economy: ``Fat tails'' with characteristic scales, European Physical Journal B 2,
525-539.

Lo, A.W. and A.C.  MacKinlay, 1999,
A Non-Random Walk down Wall Street (Princeton University Press).

Lux, L., 1996,  The stable Paretian hypothesis and the frequency of large
returns: an examination of major German stocks, Appl. Financial Economics 6, 463-475.

Mandelbrot, B.B., 1997,
Fractals and scaling in finance : discontinuity, concentration, risk,
New York : Springer.

Maslov, S. and Zhang, Y.C., 1999,
Probability distribution of drawdowns in risky investments, Physica A 262,
N1-2, 232-241.

McNeil, A.J. (1999) Extreme value theory for risk managers, preprint
ETH Zentrum Zurich.

Mood, A., 1940, The distribution theory of runs, Annals of Mathematical Statistics 11,
367-392.

M\"uller, U.A., Dacorogna, M.M., Dav\'e, R.D., Pictet, O.V., Olsen, R.B. and
Ward, J.R., 1995, preprint O\&A Research Group,
$http://www.olsen.ch/library/research/oa_working.html$

M\"uller, U.A., M.M. Dacorogna, R. Dav\'e, R.B. Olsen, O.V. Pictet and J.E. von
Weizsîcker, 1997, Volatilities of Different Time Resolutions - Analyzing the Dynamics of Market
Components, Journal of Empirical Finance 4, 213-240.

Muzy, J.-F., J. Delour and E. Bacry, 2000, Modelling fluctuations of financial
time series: from cascade process to stochastic volatility model,
European Physical Journal 17, 537-548.

Muzy, J.-F., D. Sornette, J. Delour and A.~Arneodo, 2001,
Multifractal returns and Hierarchical Portfolio Theory, 
Quantitative Finance 1 (1), 131-148.

Pagan, A., 1996, The Econometrics of Financial Markets,
Journal of Empirical Finance 3, 15 - 102.

Phoa, W. (1999) Estimating credit spread risk using extreme value theory -- Application of
actuarial disciplines to finance, Journal of Portfolio Management 25, 69-73.

Plerou, V., Gopikrishnan, P., Amaral, L.A.N., Meyer, M. \& Stanley, H.E., 1999,
Scaling of distribution of price fluctuations of individual companies, 
Physical Review E 60, 6519-6529.

Robinson, P.M., 1979, The estimation of a non-linear moving average model,
Stochastic processes and their applications 5 (February), 81-90.

Sornette, D. and R. Cont, 1997,
Convergent multiplicative processes repelled from zero: power laws and
truncated power laws, Journal of Physique I France 7, 431-444.

Sornette, D., 2000,  Critical Phenomena in Natural Sciences, 
Chaos, Fractals, Self-organization and Disorder: Concepts and Tools,
432 pp., 87 figs., 4 tabs  (Springer Series in Synergetics, Heidelberg).

Sornette, D., P. Simonetti and J. V. Andersen, 2000,
 $\phi^q$-field theory for Portfolio optimization: ``fat tails'' and
non-linear correlations, Physics Reports 335, 19-92.

Sornette, D., J. V. Andersen and P. Simonetti, 2000,
Portfolio Theory for ``Fat Tails'',
International Journal of Theoretical and Applied Finance 3 (3), 523-535.

Sornette, D. and A. Johansen, 2001,
Significance of log-periodic precursors to financial crashes,
Quantitative Finance 1 (4), 452-471.

Sornette, D. and J.V. Andersen, 2001,
Quantifying Herding During Speculative Financial Bubbles, preprint at
http://arXiv.org/abs/cond-mat/0104341

Thoman,~D.R., Bain,~L.J. and Antle,~C.E., 1970,
Maximum likelihood estimation, exact
confidence intervals for reliability, and tolerance limits in the Weibull distribution,
Technometrics 12,~363-371.

von Mises, R., 1921, Zeit. f. angew. Math. u. Mech. 1, 298.

Vries, C.G. de, 1994, Stylized Facts of Nominal Exchange
Rate Returns, S. 348 - 89 in: van der Ploeg, F., ed.,
The Handbook of International Macroeconomics. Blackwell: Oxford.

%V.S. L'vov, A. Pomyalov and I. Procaccia,
%Outliers, Extreme Events and Multiscaling, preprint of the
%Weizmann Institute of Science, available at 
%$http://babbage.sissa.it/abs/nlin.CD/0009049$

\newpage

%FIGURE 1
\begin{figure}
\begin{center}
\caption{\protect\label{djdd-dunorm} Normalized natural logarithm of the 
cumulative distribution of drawdowns and of the complementary cumulative 
distribution of drawups for the Dow Jones Industrial Average index (US stock market). 
The two continuous lines show the fits of these
two distributions with the Weibull exponential distribution defined by
formula (\protect\ref{stretched}). Negative values such as $-0.20$ and $-0.10$ correspond
to drawdowns of amplitude respectively equal to $20\%$ and $10\%$. Similarly, 
positive values corresponds to drawups with,
for instance, a number $0.2$ meaning a drawup of $+20\%$. 
}
\end{center}
\end{figure}

\clearpage

%FIGURE 2
\begin{figure}
\begin{center}
\caption{\protect\label{spdd-dunorm} Same as figure \protect\ref{djdd-dunorm}
for the S\&P500 index (US stock market).
}
\end{center}
\end{figure}

\clearpage

%FIGURE 3
\begin{figure}
\begin{center}
\caption{\protect\label{nasdd-dunorm} Same as figure \protect\ref{djdd-dunorm}
for the Nasdaq composite index (US stock market).
}
\end{center}
\end{figure}

\clearpage

%FIGURE 4
\begin{figure}
\begin{center}
\epsfig{file=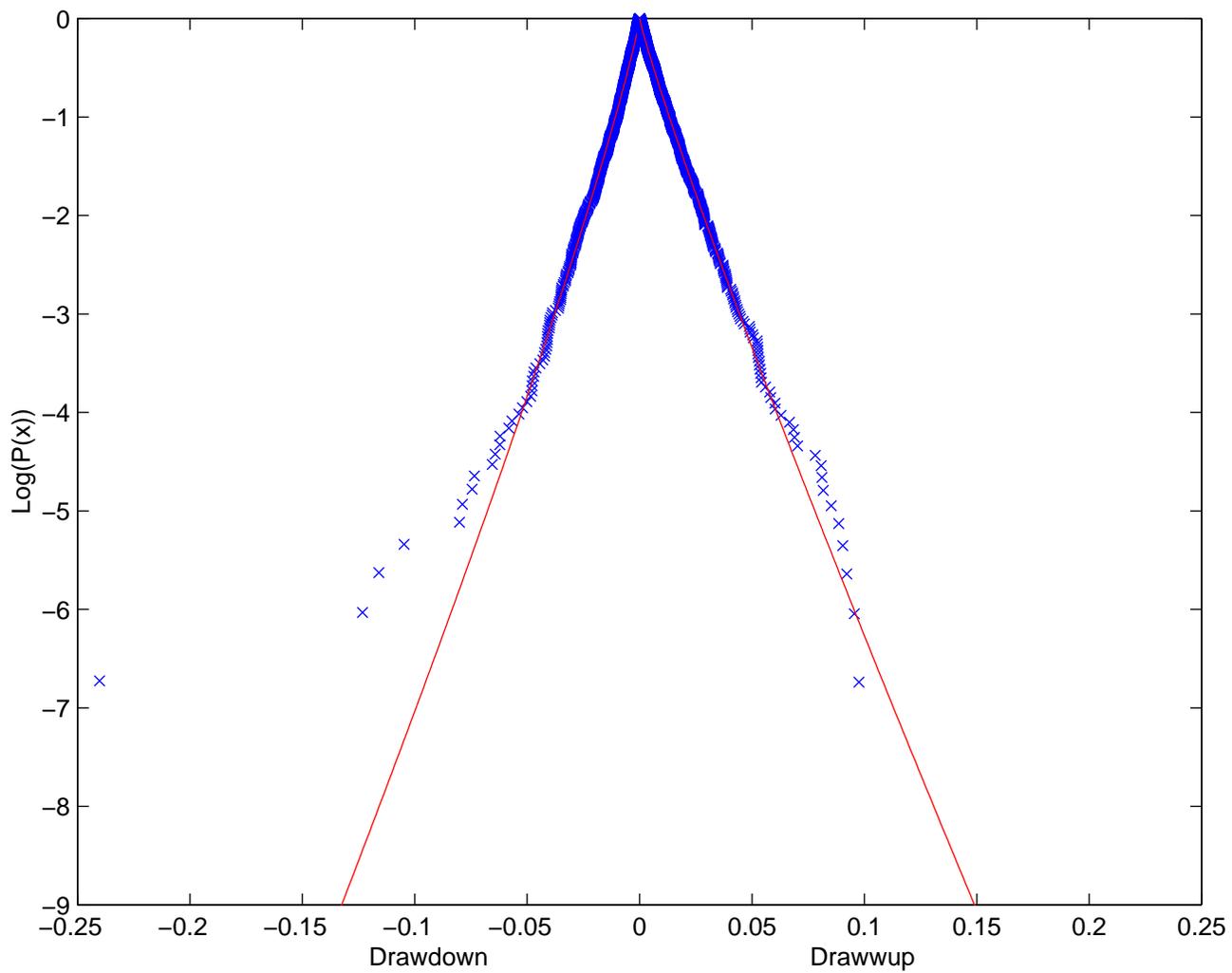}
\caption{\protect\label{candd-dunorm} Same as figure \protect\ref{djdd-dunorm}
for the TSE 300 composite index (Toronto, Canada).
}
\end{center}
\end{figure}

\clearpage

%FIGURE 5
\begin{figure}
\begin{center}
\epsfig{file=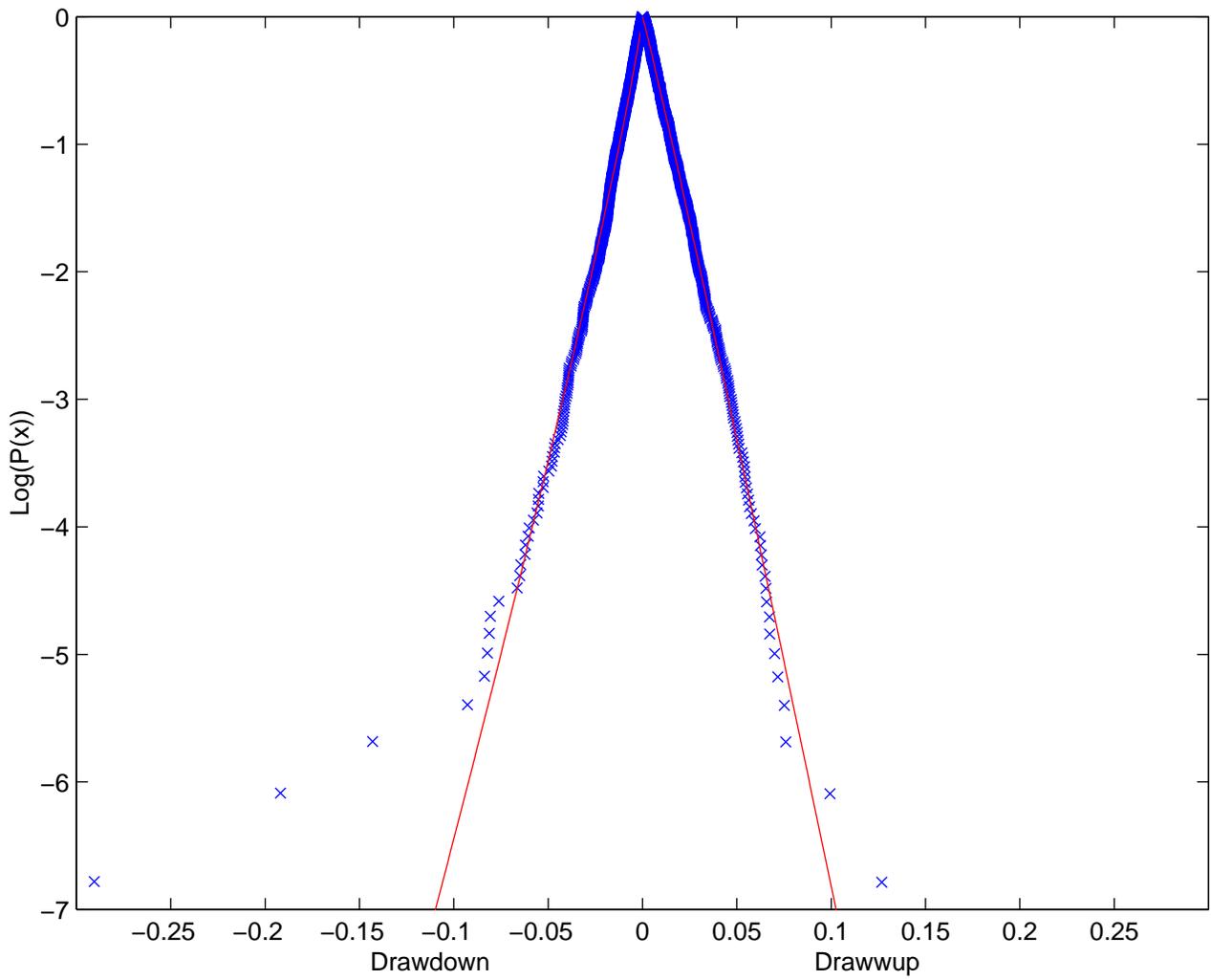}
\caption{\protect\label{austdd-dunorm} Same as figure \protect\ref{djdd-dunorm}
for the All Ordinaries index (Sydney stock exchange, Australia).
}
\end{center}
\end{figure}

\clearpage

%FIGURE 6
\begin{figure}
\begin{center}
\epsfig{file=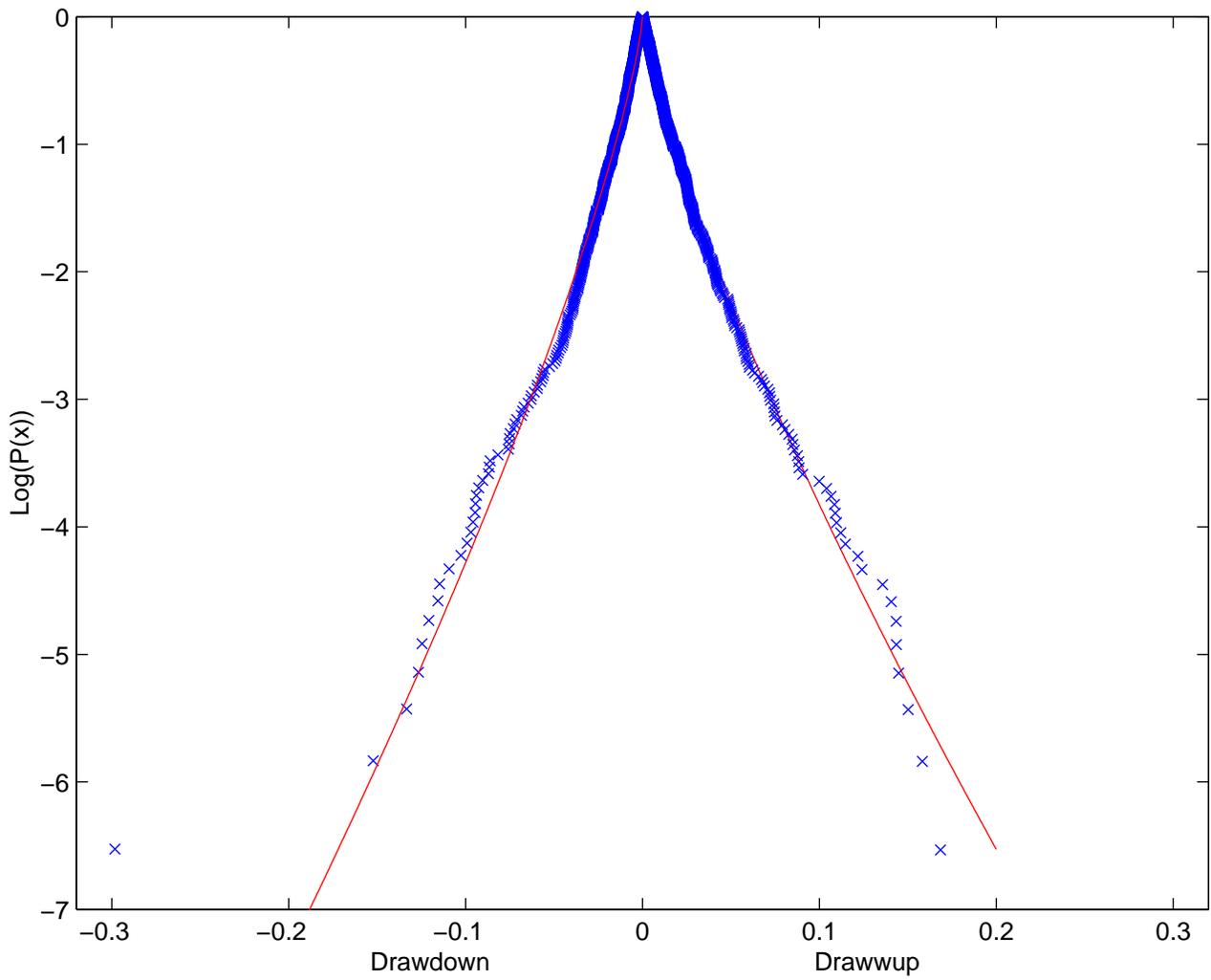}
\caption{\protect\label{singdd-dunorm} Same as figure \protect\ref{djdd-dunorm}
for the Strait Times index (Singapore stock exchange).
}
\end{center}
\end{figure}

\clearpage

%FIGURE 7
\begin{figure}
\begin{center}
\epsfig{file=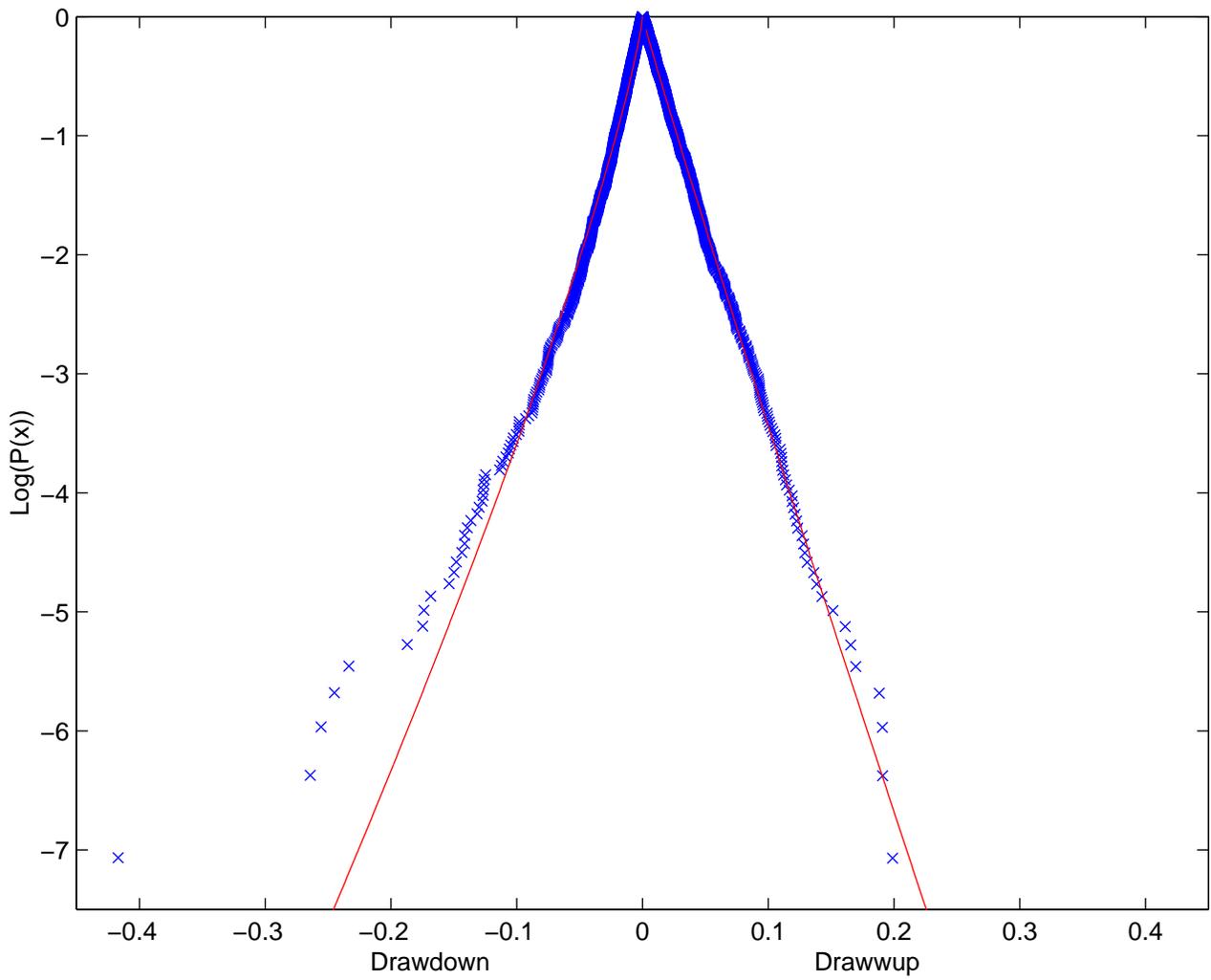}
\caption{\protect\label{hkdd-dunorm} Same as figure \protect\ref{djdd-dunorm}
for the Hang Seng index (Hong Kong stock exchange).
}
\end{center}
\end{figure}

\clearpage

%FIGURE 8
\begin{figure}
\begin{center}
\epsfig{file=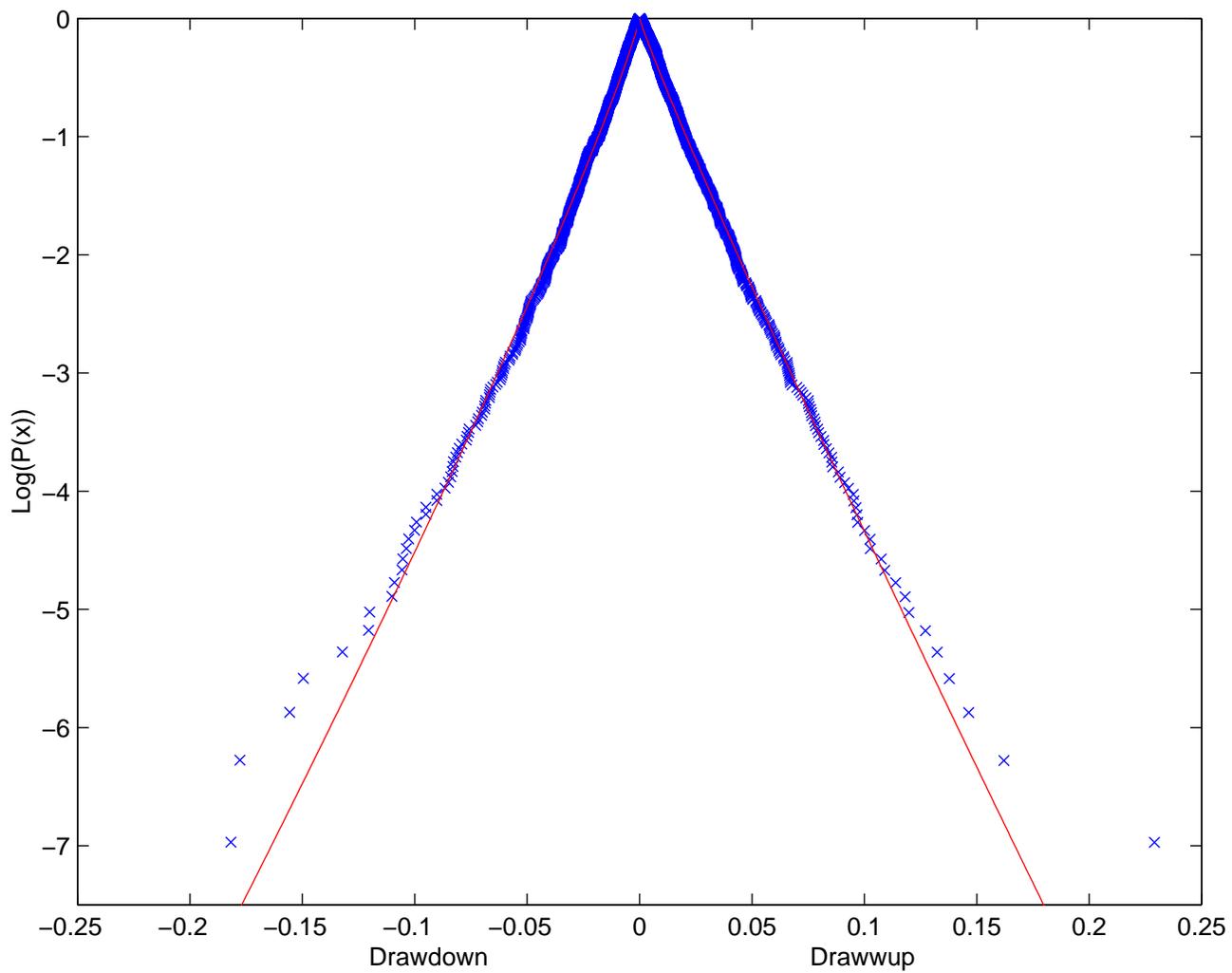}
\caption{\protect\label{nikdd-dunorm} Same as figure \protect\ref{djdd-dunorm}
for the Nikkei 225 index (Tokyo stock exchange, Japan).
}
\end{center}
\end{figure}

\clearpage

%FIGURE 9 
\begin{figure}
\begin{center}
\epsfig{file=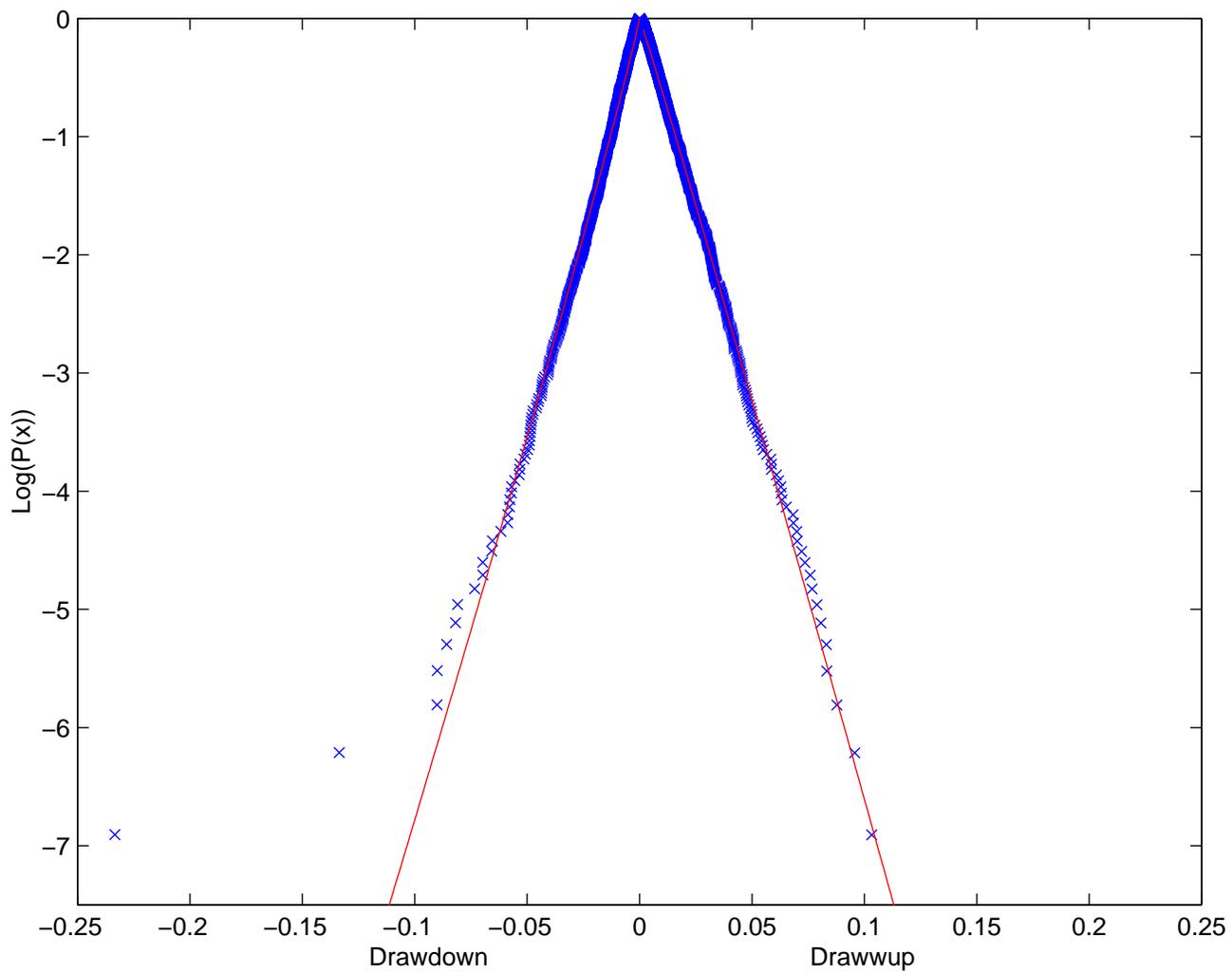}
\caption{\protect\label{ftsedd-dunorm} Same as figure \protect\ref{djdd-dunorm}
for the FTSE 100 composite index (London stock exchange, U.K.).
}
\end{center}
\end{figure}

\clearpage

%FIGURE 10 
\begin{figure}
\begin{center}
\epsfig{file=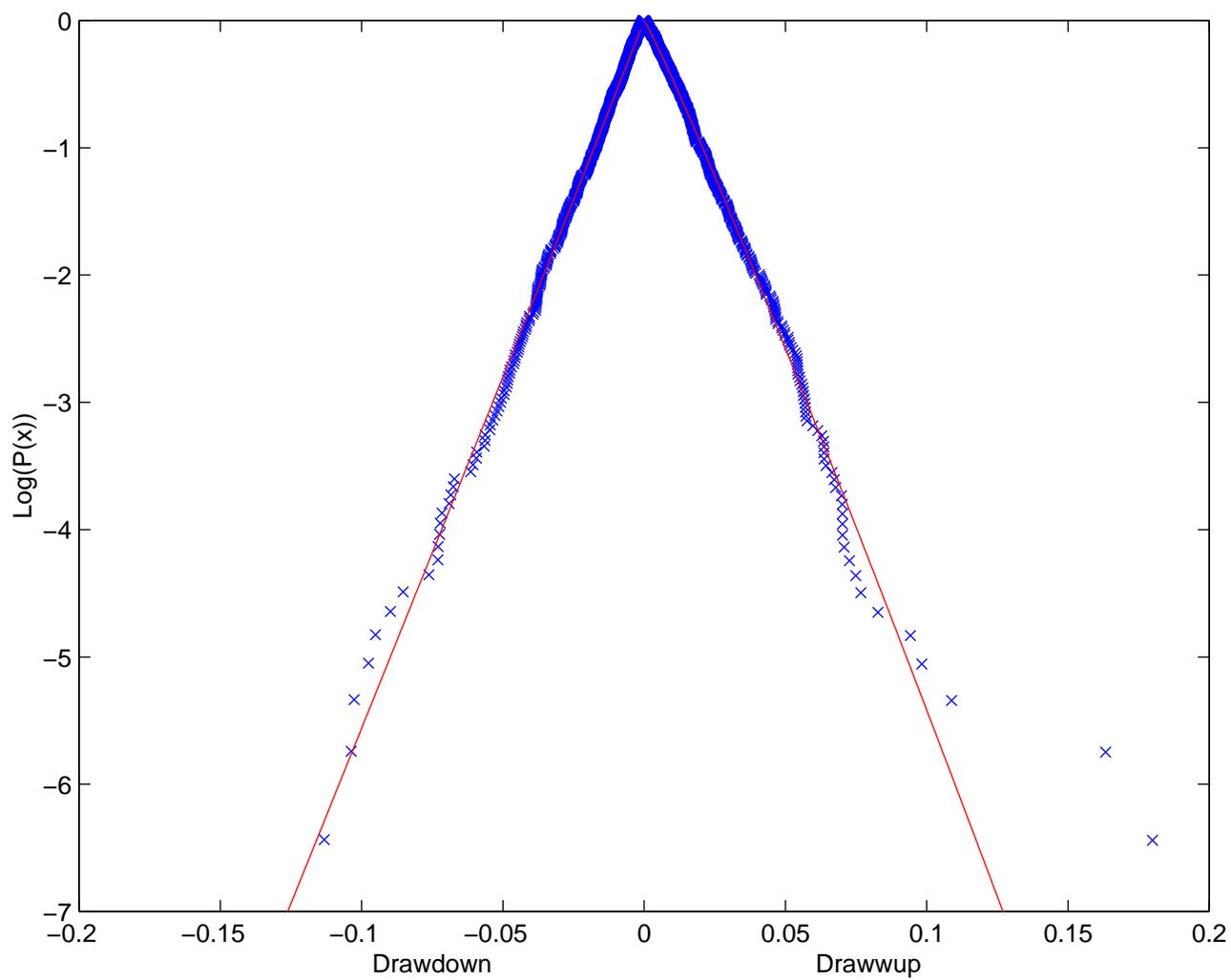}
\caption{\protect\label{cacdd-dunorm} Same as figure \protect\ref{djdd-dunorm}
for the CAC40 index (Paris stock exchange, France).
}
\end{center}
\end{figure}

\clearpage

%FIGURE 11 
\begin{figure}
\begin{center}
\epsfig{file=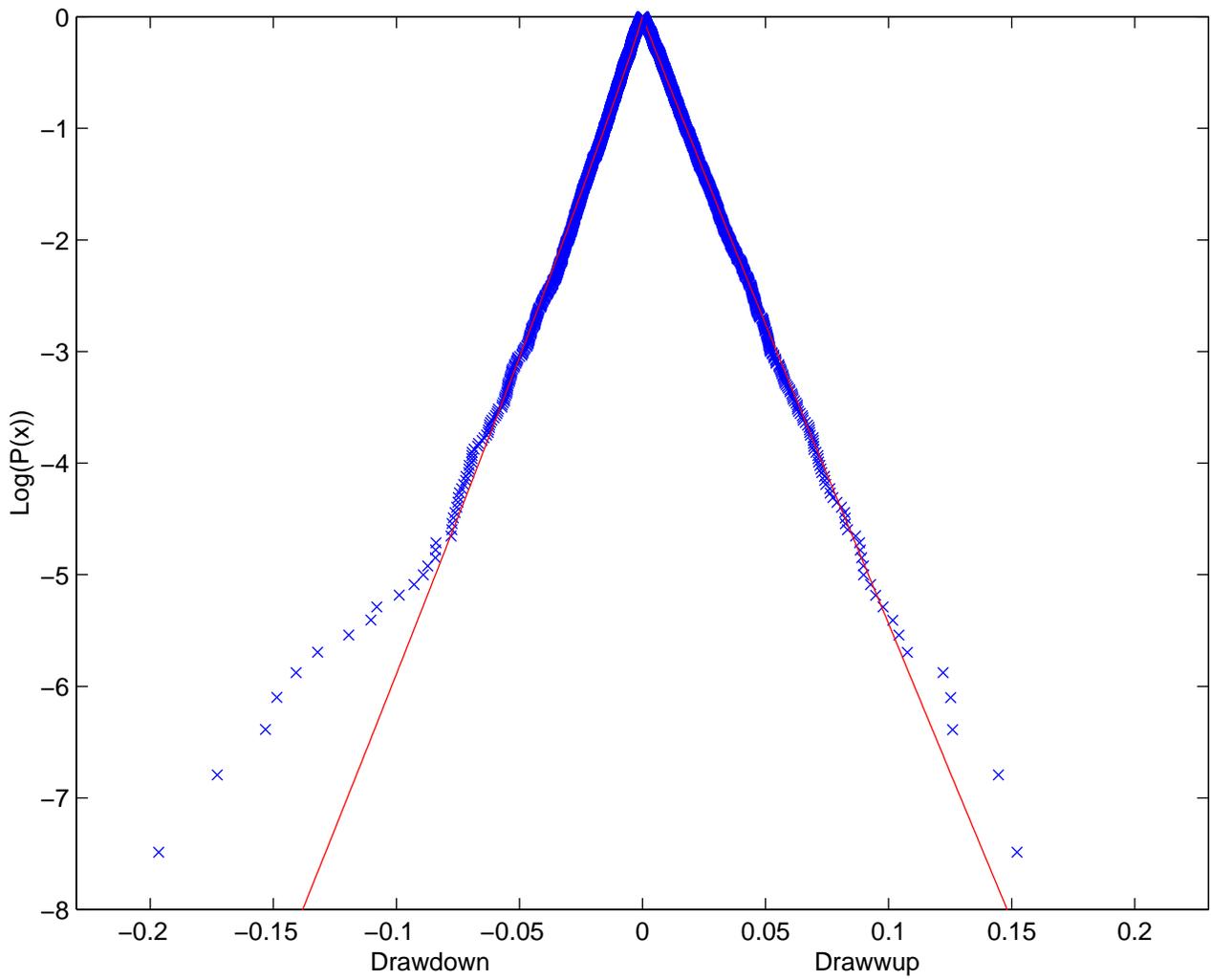}
\caption{\protect\label{daxdd-dunorm} Same as figure \protect\ref{djdd-dunorm}
for the DAX index (Frankfurt stock exchange, Germany).
}
\end{center}
\end{figure}

\clearpage

%FIGURE 12 
\begin{figure}
\begin{center}
\epsfig{file=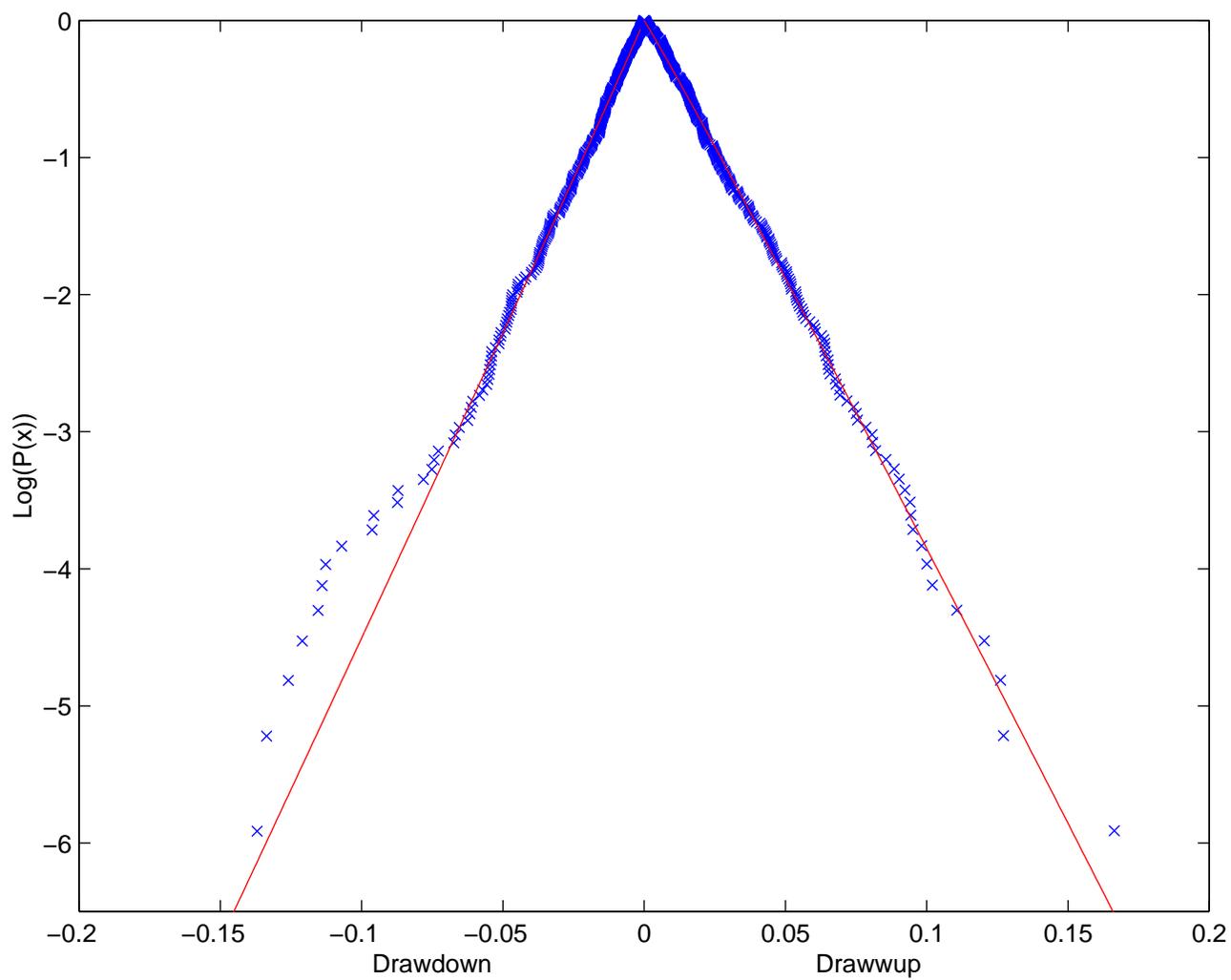}
\caption{\protect\label{itdd-dunorm} Same as figure \protect\ref{djdd-dunorm}
for the MIBTel index (Milan stock exchange, Italy).
}
\end{center}
\end{figure}

\clearpage

%FIGURE 13 
\begin{figure}
\begin{center}
\epsfig{file=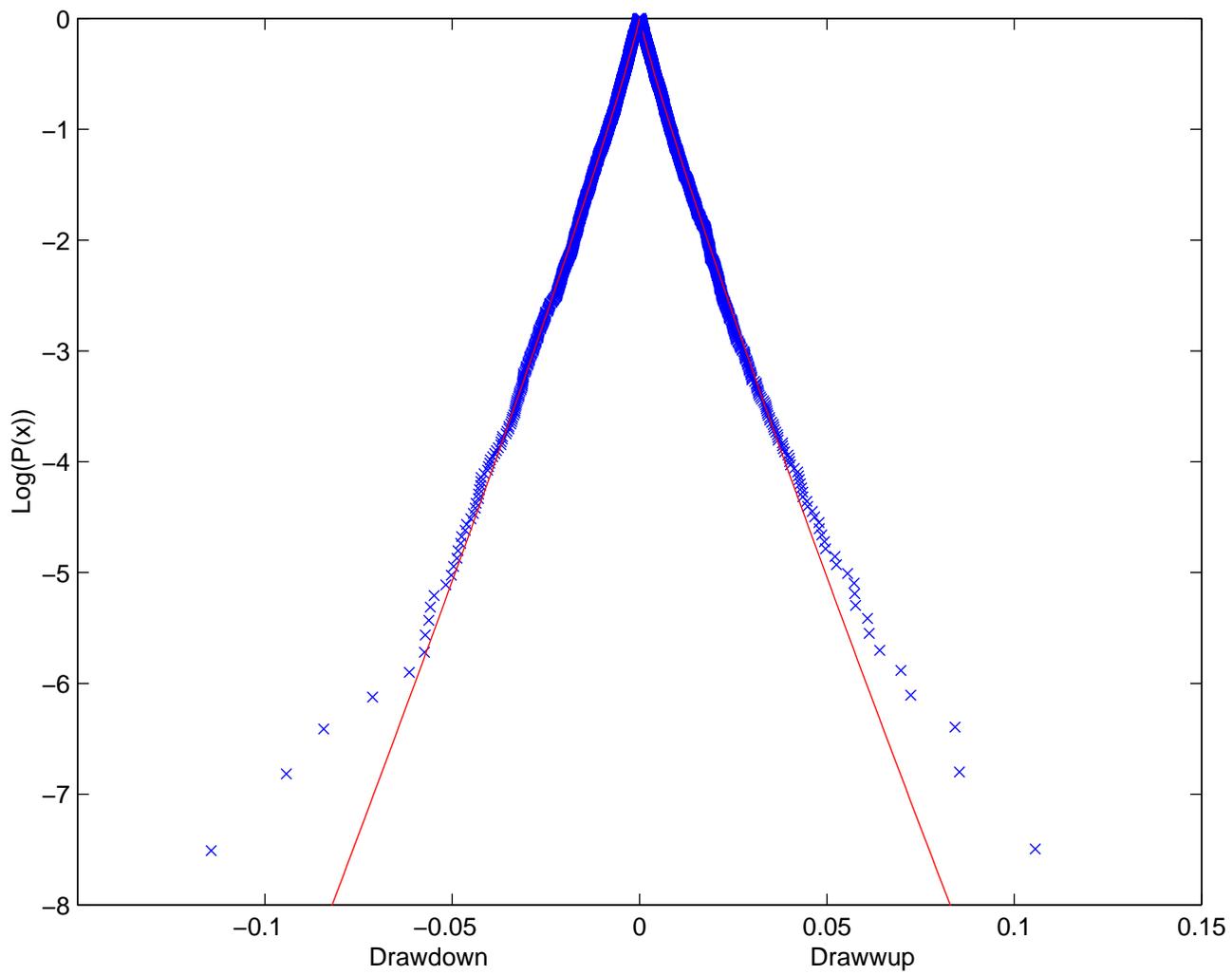}
\caption{\protect\label{usdmdd-dunorm} Same as figure \protect\ref{djdd-dunorm}
for the US\$/DM exchange rate, {\it i.e.}, US \$ expressed in Deutchmark).
}
\end{center}
\end{figure}

\clearpage

%FIGURE 14 
\begin{figure}
\begin{center}
\epsfig{file=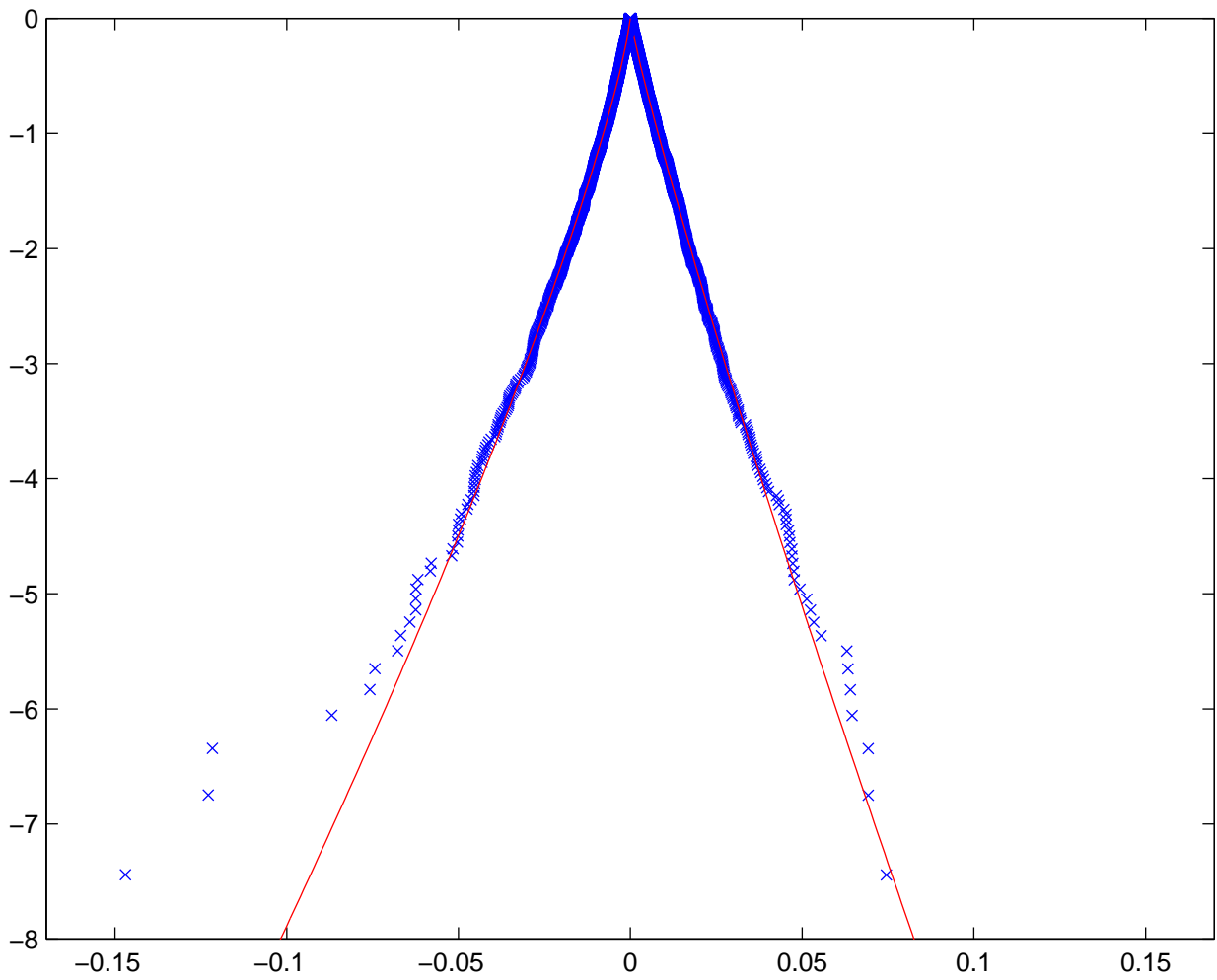}
\caption{\protect\label{usyendd-dunorm} Same as figure \protect\ref{djdd-dunorm}
for the US\$/Yen exchange rate {\it i.e.}, US \$ expressed in Japanese Yen).
}
\end{center}
\end{figure}

\clearpage

%FIGURE 15 
\begin{figure}
\begin{center}
\epsfig{file=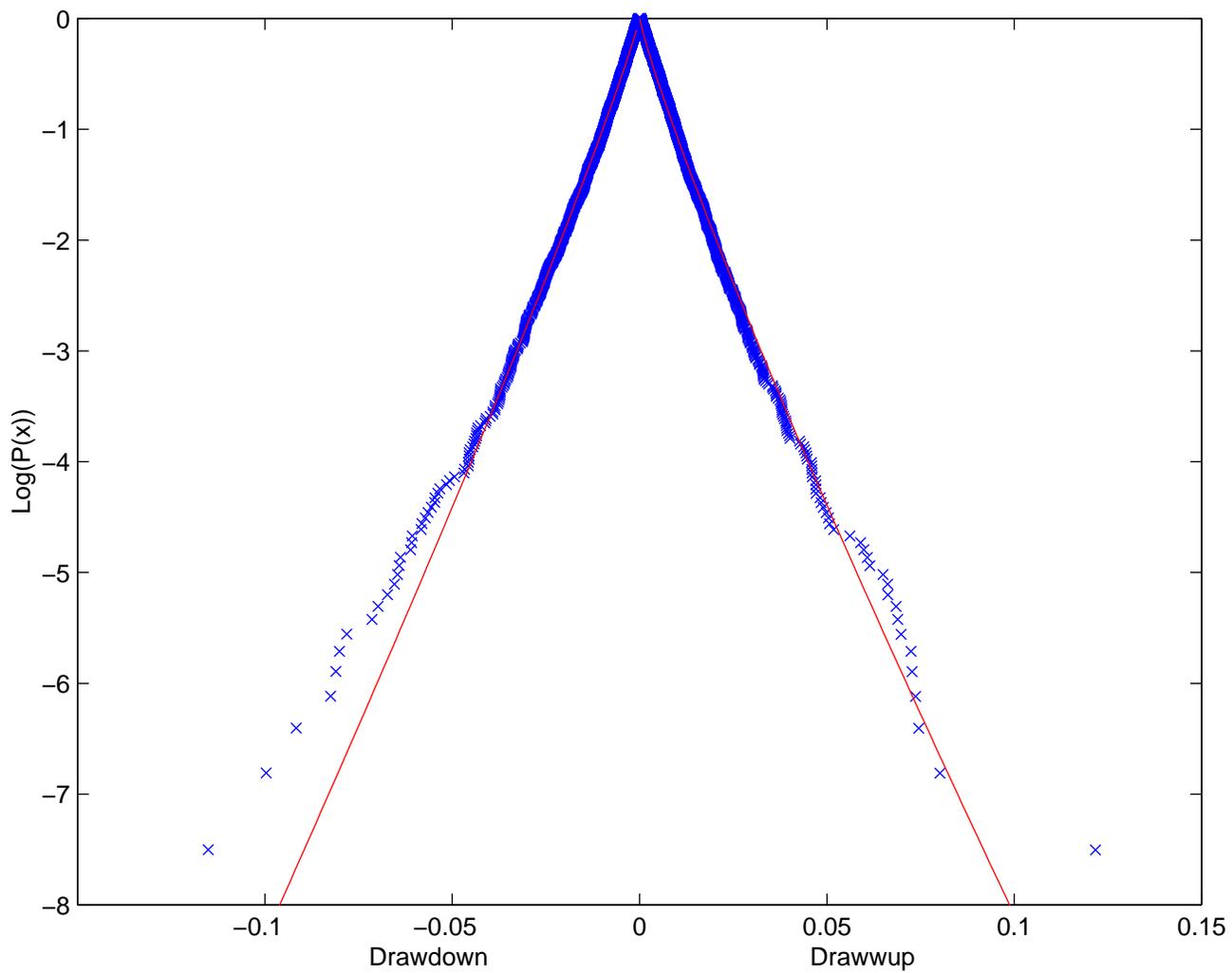}
\caption{\protect\label{uschfdd-dunorm} Same as figure \protect\ref{djdd-dunorm}
for the US\$/CHF exchange rate {\it i.e.}, US \$ expressed in Swiss franc).
}
\end{center}
\end{figure}

\clearpage

%FIGURE 16 
\begin{figure}
\begin{center}
\epsfig{file=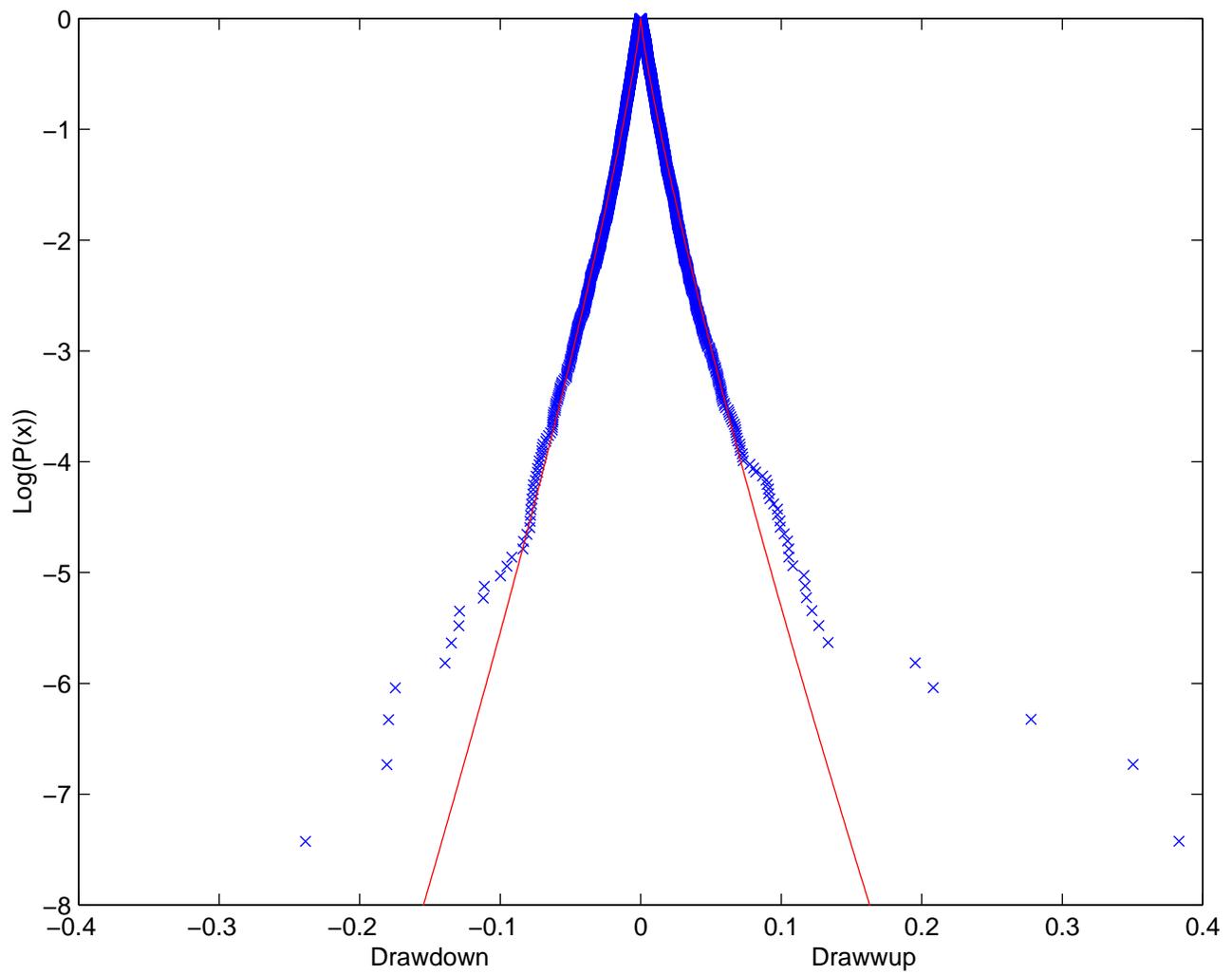}
\caption{\protect\label{audd-dunorm} Same as figure \protect\ref{djdd-dunorm}
for Gold.
}
\end{center}
\end{figure}

\clearpage

%FIGURE 17 
%\begin{figure}
%\begin{center}
%\epsfig{file=gbpusdd-dunorm.eps}
%\caption{\protect\label{gbpusdd-dunorm} Same as figure \protect\ref{djdd-dunorm}
%for the British pound expressed in US dollar.}
%\end{center}
%\end{figure}

%\clearpage

%FIGURE 18 
%\begin{figure}
%\begin{center}
%\epsfig{file=gbpdmdd-dunorm.eps}
%\caption{\protect\label{gbpdmdd-dunorm} Same as figure \protect\ref{djdd-dunorm}
%for the British pound expressed in German mark.}
%\end{center}
%\end{figure}

%\clearpage

%FIGURE 17 
\begin{figure}
\begin{center}
\epsfig{file=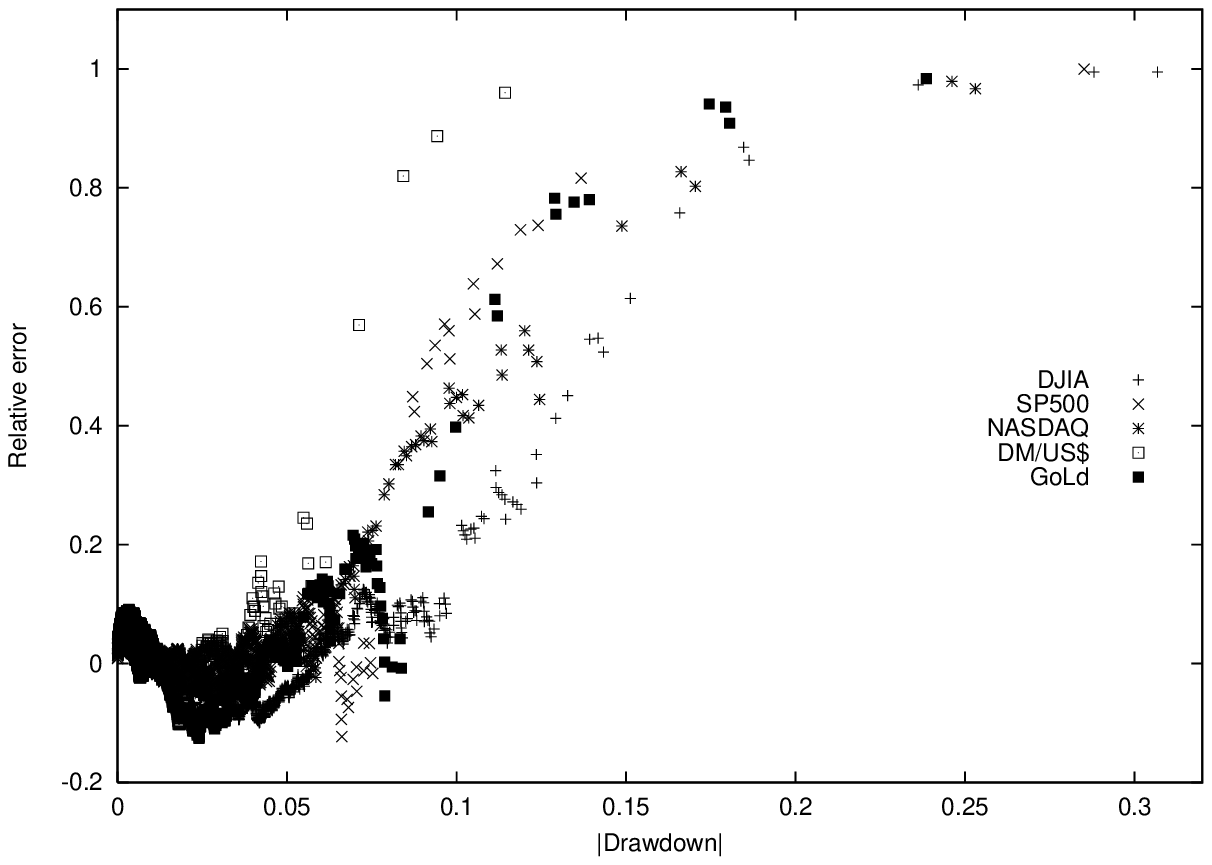}
\caption{\protect\label{alldderror} Difference (error plot) between the cumulative
distribution of drawdowns and the best fit with the stretched exponential
model (\ref{stretched}) for the DJIS, the S\&P500, the Nasdaq composite index, 
the German mark in US \$ and Gold. 
}
\end{center}
\end{figure}

\clearpage

%FIGURE 18 
\begin{figure}
\begin{center}
\epsfig{file=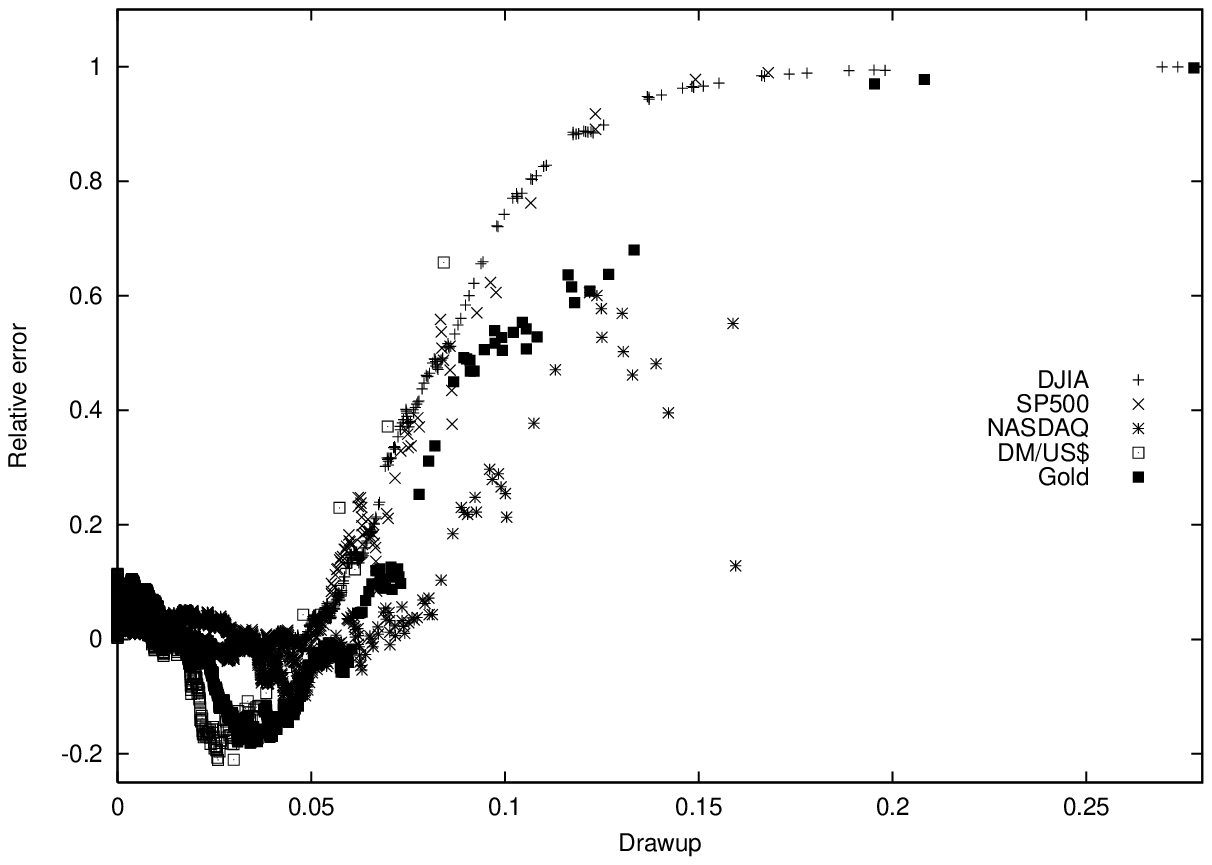}
\caption{\protect\label{alluperror} Difference (error plot) between the 
complementary cumulative
distribution of drawups and the best fit with the stretched exponential
model (\ref{stretched}) for the DJIS, the S\&P500, the Nasdaq composite index, 
the German mark in US \$ and Gold. 
}
\end{center}
\end{figure}

\clearpage

%FIGURE 19 
\begin{figure}
\begin{center}
\epsfig{file=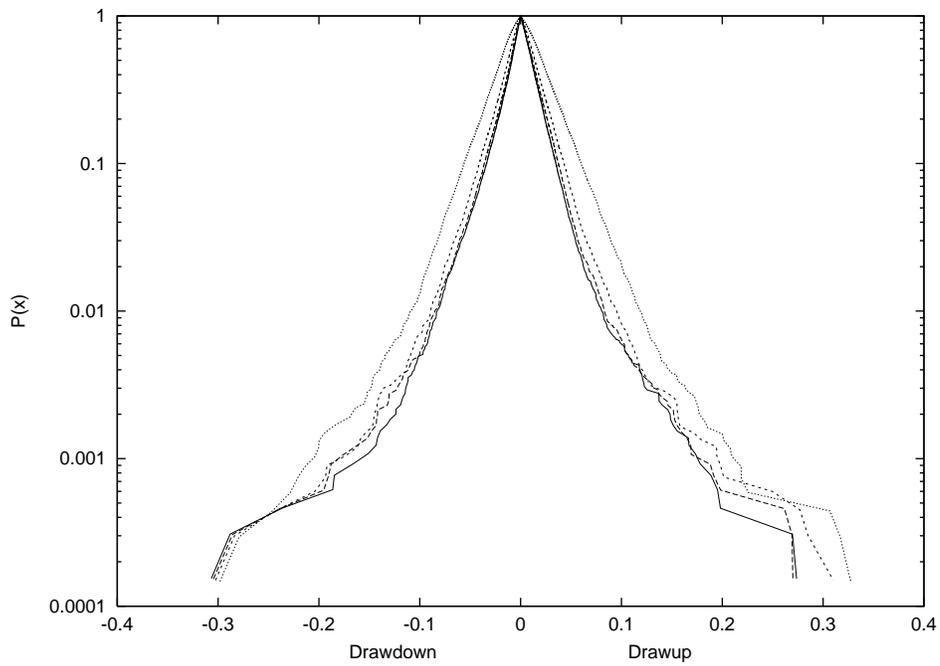}
\caption{\protect\label{djnoiseline}  Cumulative distribution of drawdowns
and complementary distributions of drawups for the
 ``time series neighborhoods'' 
defined in section \protect\ref{hghngw} obtained by adding noise
to the DJIA time series. The four curves corresponds respectively to
$A=0$ (no added noise), $A=0.5$ (added noise of standard deviation $\sigma/6$),
$A=1$ (added noise of of standard deviation $\sigma/3$) and
$A=2$ (added noise of of standard deviation $2\sigma/3$), where $\sigma$ is the
standard deviation of the returns.
}
\end{center}
\end{figure}

\clearpage

%FIGURE 20 
\begin{figure}
\begin{center}
\epsfig{file=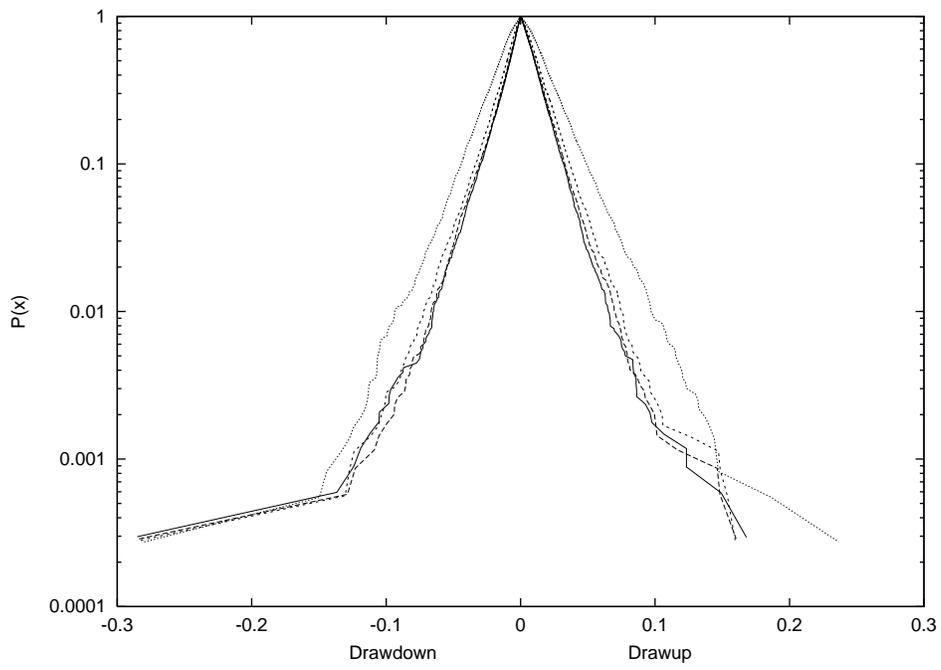}
\caption{\protect\label{spnoiseline} Same as figure \ref{djnoiseline} for
the S\&P500 index.
}
\end{center}
\end{figure}

\clearpage

%FIGURE 21 
\begin{figure}
\begin{center}
\epsfig{file=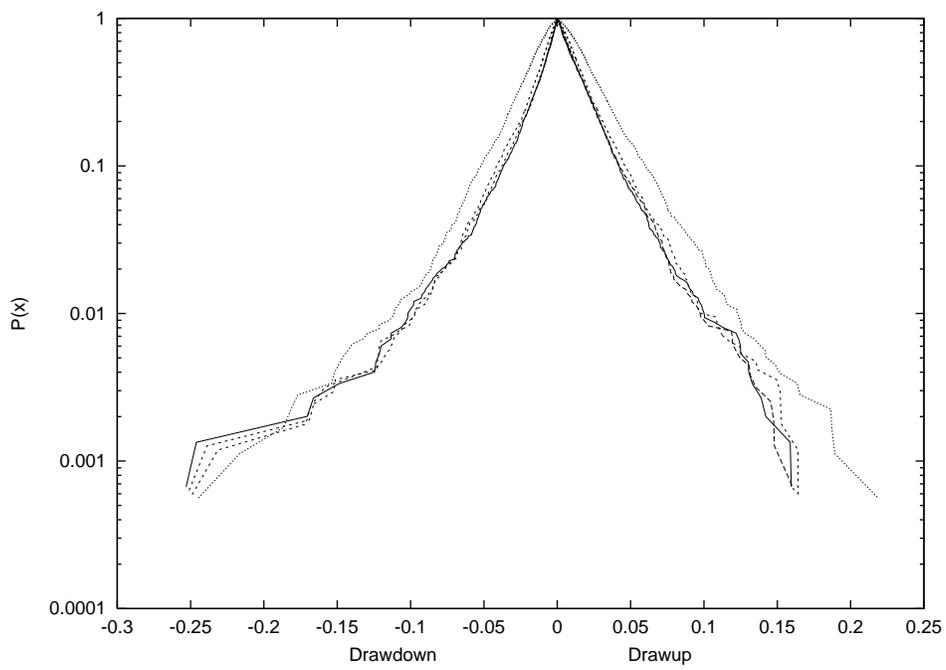}
\caption{\protect\label{nasnoiseline} Same as figure \ref{djnoiseline} for
the Nasdaq composite index.
}
\end{center}
\end{figure}

\clearpage

%FIGURE 22 
\begin{figure}
\begin{center}
\caption{\protect\label{rescaleall-1} Cumulative distribution of 
drawdowns and complementary cumulative distribution of drawups for
$29$ companies, which include the $20$ largest USA companies in terms of capitalisation
according to Forbes at the beginning of the year 2000, and in addition
Coca Cola (Forbes number 25), Qualcomm (number 30), 
Appl. Materials (number 35), Procter \& Gamble (number 38)
JDS Uniphase (number 39), General Motors (number 43), Am. Home Prod. (number 46),
Medtronic (number 50) and Ford (number 64). This figure plots each
distribution $N_c$ normalized by its corresponding factor $A$ 
as a function of the variable $y \equiv |x|/\chi)^z$, where 
 $\chi$ and $z$ are specific to each distribution and obtained
 from the fit to the stretched exponential model (\ref{stretched}).
}
\end{center}
\end{figure}

\clearpage

FIGURE 23
\begin{figure}
\begin{center}  
\parbox[l]{8.5cm}{
\epsfig{file=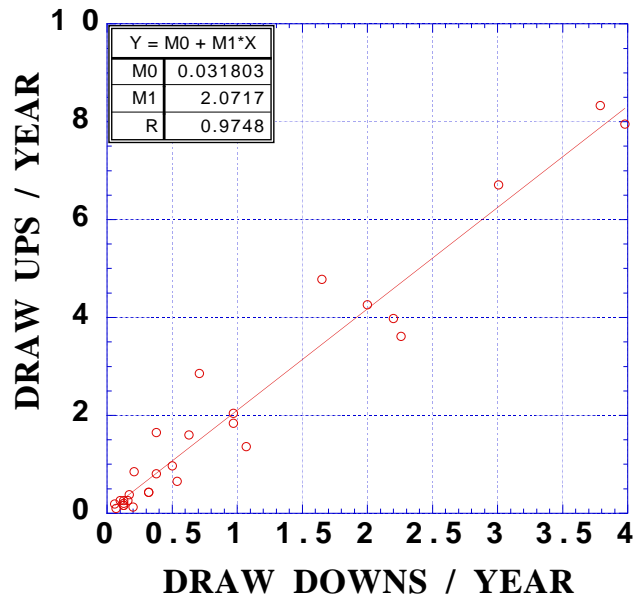,height=8cm,width=8.5cm}
\caption{\label{Du-DDregress}  Number of
drawups of amplitude larger than $15\%$
as a function of the number of drawdowns of amplitude larger 
than $15\%$ for all the companies analyzed
here and shown respectively in tables \ref{dutable} and \ref{ddtable}.
The linear regression is shown as the straight line and the parameters
are given in the table within the figure. The correlation coefficient is 
$0.62$.}}
\end{center}
\end{figure}

\clearpage

\begin{table}[]
\begin{center}
\begin{tabular}{|c|c|c|c|c|c|c|} \hline
rank & starting time & index value & duration (days) & loss  \\ \hline
1 & $87.786$ & $2508.16$ & 4 & $-30.7\%$ \\ \hline
2 & $14.579$ & $76.7$  & 2 & $-28.8\%$ \\ \hline
3 & $29.818$ & $301.22$  & 3 & $-23.6\%$ \\ \hline
4 & $33.549$ & $108.67$ & 4 & $-18.6\%$ \\ \hline
5 & $32.249$ & $77.15$ & 8 & $-18.5\%$ \\ \hline
6 & $29.852$ & $238.19$ & 4 & $-16.6\%$ \\ \hline
7 & $29.835$ & $273.51$ & 2 & $-16.6\%$ \\ \hline
8 & $32.630$ & $67.5$ & 1 & $-14.8\%$ \\ \hline
9 & $31.93$ & $90.14$ & 7 & $-14.3$ \\ \hline
10 & $32.694$ & $76.54$ & 3 & $-13.9\%$ \\ \hline
11 & $74.719$ & $674.05$ & 11 & $-13.3\%$ \\ \hline
12 & $30.444$ & $239.69$ & 4 & $-12.4\%$ \\ \hline
13 & $31.735$ & $109.86$ & 5 & $-12.9$ \\ \hline
14 & $98.649$ & $8602.65$ & 4 & $-12.4\%$ \\ \hline
\end{tabular}
\vspace{5mm}
\caption{\label{largedddj} Characteristics of the 14 largest drawdowns of
the Dow Jones Industrial Average in this century. The starting 
dates are given in decimal years.}
\end{center}
\end{table}

\begin{table}[]
\begin{center}
\begin{tabular}{|c|c|c|c|c|c|} \hline
\bf{Index}     & $A$ (fixed) & $B$ & $\chi$ (\%) & $z$    & Time period of data \\ \hline
DJ             & $6486$ & $37.7$ & $1.33 \pm 0.02$ & $0.84 \pm 0.01$ &   [1900.0:2000.5] \\ \hline
S\&P           & $3363$ & $54.8$ & $1.17 \pm 0.02$ & $0.90 \pm 0.01$ &   [1940.9:2000.5] \\ \hline
Nasdaq         & $1495$ & $31.9$ & $1.32 \pm 0.04$ & $0.80 \pm 0.02$ &   [1971.1:2000.5] \\ \hline
TSE 300 Composite& $833$& $52.7$ & $1.05 \pm 0.04$ & $0.87 \pm 0.02$ &   [1971.1:2000.5] \\ \hline
All Ordinaries & $881$  & $49.7$ & $1.24 \pm 0.05$ & $0.89 \pm 0.02$ &   [1984.6:2000.5] \\ \hline
Straits Times  & $683$  & $25.7$ & $1.56 \pm 0.08$ & $0.78 \pm 0.02$ &   [1988.0:2000.5] \\ \hline
Hang Seng      & $1171$ & $23.7$ & $2.11 \pm 0.08$ & $0.82 \pm 0.02$ &   [1980.1:2000.5] \\ \hline
Nikkei 225     & $1063$ & $35.0$ & $1.84 \pm 0.07$ & $0.89 \pm 0.02$ &   [1973.0:2000.5] \\ \hline
FTSE 100       & $997$  & $58.9$ & $1.31 \pm 0.05$ & $0.94 \pm 0.02$ &   [1984.3:2000.5] \\ \hline
CAC 40         & $623$  & $54.4$ & $1.77 \pm 0.08$ & $0.99 \pm 0.03$ &   [1990.0:2000.5] \\ \hline
DAX            & $1784$ & $52.3$ & $1.55 \pm 0.04$ & $0.95 \pm 0.02$ &   [1970.0:2000.5] \\ \hline 
MIBTel         & $370$  & $43.3$ & $2.03 \pm 0.11$ & $0.98 \pm 0.03$ &   [1993.0:1999.1] \\ \hline \hline
\bf{Currencies}& $A$ (fixed) & $B$ & $\chi$ & $z$  & Time period of data \\ \hline
German Mark    & $1826$ & $77.7$ & $0.84 \pm 0.02$ & $0.91 \pm 0.02$ &   [1971.0:1999.4] \\ \hline
Japaneese Yen  & $1706$ & $54.9$ & $1.17 \pm 0.03$ & $0.90 \pm 0.02$ &   [1972.0:1999.4] \\ \hline
Swiss Franc    & $1813$ & $67.8$ & $0.97 \pm 0.03$ & $0.91 \pm 0.02$ &   [1971.0:1999.4] \\ \hline \hline
\bf{Gold}      & $1681$ & $38.6$ & $1.29 \pm 0.04$ & $0.84 \pm 0.02$ &   [1975.0:1999.8] \\ \hline
\end{tabular}
\end{center}
\caption{\label{ddindex} Parameter values obtained by fitting equation 
\protect(\ref{stretched}) to the cumulative distribution $N_c$ of drawdowns. 
In order to stabilize the fit, it has been performed as $\log(N_c) = \log(A) - 
B|x|^z$, where $A$ is the total number of drawdowns and hence fixed equivalent 
to a normalisation of the corresponding probability distribution. 
The characteristic scale $\chi$ is defined by $\chi= 1/B^{1/z}$. DJ is the 
Dow Jones Industrial Average, S\&P is the Standard and Poor 500 Index, Nasdaq 
is the Nasdaq Composite, TSE 300 Composite is the index of the stock exchange 
of Toronto, Canada, All Ordinaries is that of Sydney stock exchange, Australia,
Straits Times is that of Singapore stock exchange, Hang Seng is that of Hong Kong
stock exchange, Nikkei 225 is that of Tokyo stock exchange, Japan, FTSE 100 is 
that of London stock exchange, U.K., CAC 40 is that of Paris stock exchange, 
France, Dax is that of Frankfurt stock exchange, Germany and
MIBTel is that of Milan stock exchange. The error bars reported for $\chi$ and $z$ 
are obtained from the formulas (\ref{nbvbvnx}) and (\ref{nvbklkz}).}
\end{table}

\begin{table}[]
\begin{center}
\begin{tabular}{|c|c|c|c|c|c|}\hline
\bf{Index}     & $A$ (fixed) & $B$ & $\chi (\%)$ & $z$  & Time period of data \\ \hline
DJ I            & $6508$ & $63.8$ & $1.50 \pm 0.02$   & $0.99 \pm 0.01$ &   [1900.0:2000.5] \\ \hline
DJ II    & $\ln(A)\approx 8.09$ & $16.5$ & $0.73 \pm 0.02$   & $0.57 \pm 0.005$ &   [1900.0:2000.5] \\ \hline
S\&P           & $3394$ & $79.7$ & $1.43 \pm 0.04$    & $1.03 \pm 0.02$ &   [1940.9:2000.5] \\ \hline
Nasdaq         & $1495$ & $38.9$ & $1.71 \pm 0.05$     & $0.90 \pm 0.02$ &   [1971.1:2000.5] \\ \hline
TSE 300 Composite& $844$& $50.7$ & $1.34 \pm 0.05$    & $0.91 \pm 0.02$ &   [1984.3:2000.5] \\ \hline
All Ordinaries & $885$  & $74.9$ & $1.58 \pm 0.05$    & $1.04 \pm 0.03$ &   [1984.6:2000.5] \\ \hline
Straits Times  & $687$  & $22.6$ & $1.74 \pm 0.09$    & $0.77 \pm 0.02$ &   [1988.0:2000.5] \\ \hline
Hang Seng      & $1175$ & $31.3$ & $2.59 \pm 0.08$   & $0.96 \pm 0.02$ &   [1980.1:2000.5] \\ \hline
Nikkei 225     & $1066$ & $37.2$ & $2.05 \pm 0.07$     & $0.93 \pm 0.02$ &   [1973.0:2000.5] \\ \hline
FTSE 100       & $999$  & $71.1$ & $1.59 \pm 0.05$   & $1.03 \pm 0.03$ &   [1984.3:2000.5] \\ \hline
CAC 40         & $627$  & $64.0$ & $2.05 \pm 0.08$    & $1.07 \pm 0.03$ &   [1990.0:2000.5] \\ \hline
DAX            & $1785$ & $52.1$ & $1.77 \pm 0.05$   & $0.98 \pm 0.02$ &   [1970.0:2000.5] \\ \hline 
MIBTel         & $369$  & $41.6$ & $2.57 \pm 0.12$   & $1.03 \pm 0.04$ &   [1993.0:1999.1] \\ \hline \hline
\bf{Currencies}&  $A$ (fixed) & $B$ & $\chi (\%)$ & $z$ & Time period of data \\ \hline
German Mark    & $1797$ & $63.0$ & $0.81 \pm 0.02$    & $0.86 \pm 0.02$ &   [1971.0:1999.39] \\ \hline
Japaneese Yen  & $1710$ & $74.6$ & $0.79 \pm 0.02$    & $0.89 \pm 0.02$ &   [1972.0:1999.39] \\ \hline
Swiss Franc    & $1815$ & $61.5$ & $0.93 \pm 0.03$     & $0.88 \pm 0.02$ &   [1971.0:1999.39] \\ \hline \hline
\bf{Gold}      & $1676$ & $36.7$ & $1.37 \pm 0.04$   & $0.84 \pm 0.02$ &   [1975.0:1999.8] \\ \hline
\end{tabular} 
\end{center}
\caption{\label{duindex}Parameter values obtained by fitting equation 
(\protect\ref{stretched}) to the cumulative distribution $N_c$ drawups. In order 
to stabilize the fit, it has been performed as $\log(N_c) = \log(A) - B|x|^z$, where 
$A$ is the total number of drawups and hence fixed equivalent to a normalisation 
of the corresponding probability distribution. DJ is the Dow Jones Industrial 
Average, S\&P is the Standard and Poor 500 Index, Nasdaq is the Nasdaq Composite, 
TSE 300 Composite is the index of the stock exchange of Toronto, Canada, All 
Ordinaries is that of Sydney stock exchange, Australia, Straits Times is that of 
Singapore stock exchange, Hang Seng is that of Hong Kong stock exchange, Nikkei 
225 is that of Tokyo stock exchange, Japan, FTSE 100 is that of London stock 
exchange, U.K., CAC 40 is that of Paris stock exchange, France, Dax is that of 
Frankfurt stock exchange, Germany and MIBTel is that of Milan stock exchange.}
\end{table}

\begin{table}[]
\begin{center}
\begin{tabular}{|c|c|c|c|c|c|c|} \hline
{\bf Company}    & $A$ (fixed) & $B$ & $\chi (\%)$   & $z$    & Time period     & DD/year $> 15\%$ \\ \hline
MicroSoft  & $884$       & $36.1$ &  $3.18 \pm 0.11$  & $1.04 \pm 0.03$ & [1986.2:2000.5] & $0.63$   \\ \hline
Cisco      & $620$       & $41.4$ &  $4.04 \pm 0.15$  & $1.16 \pm 0.04$ & [1990.2:2000.5] & $0.97$  \\ \hline
General Elec.& $1866$    & $51.6$ &  $2.09 \pm 0.05$  & $1.02 \pm 0.02$ & [1970.0:2000.5] & $0.13$  \\ \hline
Intel      & $846$       & $33.0$ &  $3.69 \pm 0.13$  & $1.06 \pm 0.03$ & [1986.5:2000.5] & $1.07$  \\ \hline
Exxon-Mobil& $1875$      & $70.5$ &  $1.94 \pm 0.04$  & $1.08 \pm 0.02$ & [1970.0:2000.5] & $0.07$  \\ \hline
Oracle     & $757$       & $28.7$ &  $4.73 \pm 0.16$  & $1.10 \pm 0.03$ & [1988.2:2000.5] & $2.20$  \\ \hline
Lucent     & $764$       & $40.9$ &  $3.86 \pm 0.13$  & $1.14 \pm 0.03$ & [1996.3:2000.5] & $0.71$  \\ \hline
Wall Mart  & $1643$      & $32.8$ &  $2.84 \pm 0.08$  & $0.98 \pm 0.02$ & [1972.7:2000.5] & $0.50$  \\ \hline
IBM        & $2404$      & $44.2$ &  $2.09 \pm 0.05$  & $0.98 \pm 0.02$ & [1962.0:2000.5] & $0.13$  \\ \hline
AT\&T      & $1918$      & $46.6$ &  $1.91 \pm 0.05$  & $0.97 \pm 0.02$ & [1970.0:2000.5] & $0.13$  \\ \hline
Citigroup  & $1444$      & $36.6$ &  $2.54 \pm 0.07$  & $0.98 \pm 0.02$ & [1977.0:2000.5] & $0.21$  \\ \hline
Sun Microsystem & $832$  & $30.1$ &  $4.53 \pm 0.15$ & $1.10 \pm 0.03$ & [1987.2:2000.5] & $2.26$  \\ \hline
Texas Instruments& $1114$& $30.4$ &  $3.63 \pm 0.11$  & $1.03 \pm 0.02$ & [1982.0:2000.5] & $0.97$  \\ \hline
HP         & $1454$      & $39.5$ &  $3.12 \pm 0.08$  & $1.06 \pm 0.02$ & [1977.0:2000.5] & $0.38$  \\ \hline
SBC Communications& $980$& $48.1$ &  $2.08 \pm 0.07$  & $1.00 \pm 0.02$ & [1984.6:2000.5] & $0.06$  \\ \hline
Merck      & $1825$      & $48.9$ &  $2.29 \pm 0.05$  & $1.03 \pm 0.02$ & [1970.0:2000.5] & $0.13$  \\ \hline
EMC        & $721$       & $30.6$ &  $4.59 \pm 0.16$  & $1.11 \pm 0.03$ & [1989.0:2000.5] & $1.65$  \\ \hline
Pfizer     & $1144$      & $47.9$ &  $2.60 \pm 0.08$  & $1.06 \pm 0.02$ & [1982.0:2000.5] & $0.32$  \\ \hline
AOL        & $504$       & $18.1$ &  $5.37 \pm 0.25$  & $0.99 \pm 0.03$ & [1992.2:2000.5] & $3.98$  \\ \hline
MCI WorldCom& $654$      & $41.0$ &  $3.52 \pm 0.13$  & $1.11 \pm 0.03$ & [1990.2:2000.5] & $0.38$  \\ \hline   \hline
Coca Cola  & $1839$      & $42.6$ &  $2.17 \pm 0.05$  & $0.98 \pm 0.02$ & [1970.0:2000.5] & $0.16$  \\ \hline
Oualcomm   & $537$       & $26.7$ &  $5.61 \pm 0.22$  & $1.14 \pm 0.04$ & [1992.0:2000.5] & $3.01$  \\ \hline
Appl. Materials& $963$   & $30.0$ &  $5.06 \pm 0.15$  & $1.14 \pm 0.03$ & [1985.0:2000.5] & $2.00$  \\ \hline
Procter\&Gamble& $1857$  & $42.8$ &  $1.92 \pm 0.05$  & $0.95 \pm 0.02$ & [1970.0:2000.5] & $0.20$  \\ \hline
JDS Uniphase& $402$      & $21.0$ &  $5.66 \pm 0.28$  & $1.06 \pm 0.04$ & [1993.9:2000.5] & $3.79$  \\ \hline
General Motors & $1871$  & $62.3$ &  $2.50 \pm 0.05$  & $1.12 \pm 0.02$ & [1970.0:2000.5] & $0.10$  \\ \hline
Am. Home Prod.& $1100$   & $40.3$ &  $2.21 \pm 0.07$  & $0.97 \pm 0.02$ & [1982.0:2000.5] & $0.32$  \\ \hline
Medtronic  & $1072$      & $32.2$ &  $3.55 \pm 0.11$  & $1.04 \pm 0.02$ & [1982.0:2000.5] & $0.54$  \\ \hline
Ford       & $1481$      & $49.6$ &  $2.60 \pm 0.07$  & $1.07 \pm 0.02$ & [1977.0:2000.5] & $0.17$  \\ \hline
\end{tabular}
\end{center}
\caption{\label{ddtable}The parameter values obtained by fitting equation 
(\protect\ref{stretched}) to the cumulative distribution $N_c$ of drawdowns. 
In order to stabilise the fit, it has been performed as $\log(N_c) = \log(A) 
- B|x|^z$, where $A$ is the total number of drawdowns and hence fixed
equivalent to a normalisation of the corresponding probability distribution. The 
companies are the top 20 in terms of market value, number 25, number 30,
number 35, number 39 (no data for number 40 could be obtained), number 46 
(no data for number 45 could be obtained) and number 50. Three more companies 
have been added in order to get longer time series as well as representatives 
for the automobil sector. These are Procter \& Gamble (number 38), General 
Motors (number 43) and Ford (number 64). The ranking is Forbes of year 2000.}
\end{table}

\begin{table}[]
\begin{center}
\begin{tabular}{|c|c|c|c|c|c|c|}\hline
{\bf Company}    & $A$ (fixed) & $B$ & $\chi (\%)$  & $z$    & Time period     & DU/year $> 15\%$  \\ \hline
MicroSoft  & $880$      & $25.3$ &  $4.08 \pm 0.14$ & $1.01 \pm 0.03$ & [1986.2:2000.5] & $1.60$  \\ \hline
Cisco      & $617$      & $32.0$ &  $4.84 \pm 0.20$    & $1.22 \pm 0.04$ & [1990.2:2000.5] & $2.04$  \\ \hline
General Electric &$1877$& $43.9$ &  $2.45 \pm 0.06$    & $1.02 \pm 0.02$ & [1970.0:2000.5] & $0.20$  \\ \hline
Intel      & $850$      & $38.9$ &  $4.85 \pm 0.14$    & $1.21 \pm 0.03$ & [1986.5:2000.5] & $1.36$  \\ \hline
Exxon-Mobil& $1895$     & $57.8$ &  $2.18 \pm 0.05$    & $1.06 \pm 0.02$ & [1970.0:2000.5] & $0.10$  \\ \hline
Oracle     & $761$      & $25.1$ &  $5.77 \pm 0.19$    & $1.13 \pm 0.03$ & [1988.2:2000.5] & $3.98$  \\ \hline
Lucent     & $260$      & $27.4$ &  $5.20 \pm 0.30$    & $1.12 \pm 0.05$ & [1996.3:2000.5] & $2.86$  \\ \hline
Wall Mart  & $1664$     & $27.6$ &  $3.50 \pm 0.09$    & $0.99 \pm 0.02$ & [1972.7:2000.5] & $0.97$  \\ \hline
IBM        & $2400$     & $39.0$ &  $2.29 \pm 0.05$    & $0.97 \pm 0.02$ & [1962.0:2000.5] & $0.26$  \\ \hline
AT\&T      & $1854$     & $32.1$ &  $1.94 \pm 0.05$    & $0.88 \pm 0.02$ & [1970.0:2000.5] & $0.16$  \\ \hline
Citigroup  & $1423$     & $25.6$ &  $2.95 \pm 0.09$    & $0.92 \pm 0.02$ & [1977.0:2000.5] & $0.85$  \\ \hline
Sun Microsystem & $827$  & $26.4$ & $5.66 \pm 0.18$    & $1.14 \pm 0.03$ & [1987.2:2000.5] & $3.61$  \\ \hline
Texas Instruments& $1112$& $26.6$ & $4.40 \pm 0.13$    & $1.05 \pm 0.02$ & [1982.0:2000.5] & $1.84$  \\ \hline
HP         & $1447$     & $35.5$ &  $3.67 \pm 0.09$    & $1.08 \pm 0.02$ & [1977.0:2000.5] & $0.81$  \\ \hline
SBC Communications&$972$& $38.1$ &  $2.44 \pm 0.08$     & $0.98 \pm 0.02$ & [1984.6:2000.5] & $0.19$  \\ \hline
Merck      & $1816$     & $41.2$ &  $2.61 \pm 0.06$    & $1.02 \pm 0.02$ & [1970.0:2000.5] & $0.16$  \\ \hline
EMC        & $702$      & $19.4$ &  $6.26 \pm 0.23$    & $1.07 \pm 0.03$ & [1989.0:2000.5] & $4.78$  \\ \hline
Pfizer     & $1137$     & $38.1$ &  $3.12 \pm 0.09$    & $1.05 \pm 0.02$ & [1982.0:2000.5] & $0.43$  \\ \hline
AOL        & $493$      & $18.1$ &  $5.35 \pm 0.26$    & $0.99 \pm 0.03$ & [1992.2:2000.5] & $7.95$  \\ \hline
MCI WorldCom& $653$     &$25.5$  &  $4.44 \pm 0.18$    & $1.04 \pm 0.03$ & [1990.2:2000.5] & $1.65$  \\ \hline  \hline
Coca Cola  & $1841$     & $34.9$ &  $2.57 \pm 0.06$    & $0.97 \pm 0.02$ & [1970.0:2000.5] & $0.26$  \\ \hline
Oualcomm   & $537$      & $18.1$ &  $6.68 \pm 0.28$    & $1.07 \pm 0.04$ & [1992.0:2000.5] & $6.71$  \\ \hline
Appl. Materials& $961$  & $30.0$ &  $5.06 \pm 0.15$     & $1.14 \pm 0.03$ & [1985.0:2000.5] & $4.26$  \\ \hline
Procter\&Gamble& $1857$ & $47.3$ &  $2.37 \pm 0.06$    & $1.03 \pm 0.02$ & [1970.0:2000.5] & $0.13$  \\ \hline
JDS Uniphase& $419$     & $14.2$ &  $7.61 \pm 0.38$   & $1.03 \pm 0.04$ & [1993.9:2000.5] & $8.33$  \\ \hline
General Motors & $1856$ & $39.6$ &  $2.62 \pm 0.06$    & $1.01 \pm 0.02$ & [1970.0:2000.5] & $0.26$  \\ \hline
Am. Home Prod.& $1092$  & $31.5$ &  $2.55 \pm 0.09$    & $0.94 \pm 0.02$ & [1982.0:2000.5] & $0.43$  \\ \hline
Medtronic  & $1098$     & $32.2$ &  $2.49 \pm 0.08$    & $1.04 \pm 0.02$ & [1982.0:2000.5] & $0.65$  \\ \hline
Ford       & $1481$     & $35.9$ &  $2.89 \pm 0.08$    & $1.01 \pm 0.02$ & [1977.0:2000.5] & $0.38$  \\ \hline
\end{tabular}
\end{center}
\caption{\label{dutable}The parameter values obtained by fitting equation 
(\protect\ref{stretched}) to the cumulative distribution $N_c$ of drawups. In 
order to stabilise the fit, it has been performed as $\log(N_c) = \log(A) - 
B|x|^z$, where $A$ is the total number of drawups and hence fixed
equivalent to a normalisation of the corresponding probability distribution. 
The characteristic scale $\chi$ is defined by $\chi= 1/B^{1/z}$. The 
companies are the top 20 in terms of market value, number 25, number 30,
number 35, number 39 (no data for number 40 could be obtained), number 46 
(no data for number 45 could be obtained) and number 50. Three more companies 
have been added in order to get longer time series as well as representatives 
for the automobil sector. These are Procter \& Gamble (number 38), General 
Motors (number 43) and Ford (number 64). The ranking is Forbes of year 2000. 
The error bars reported for $\chi$ and $z$ 
are obtained from the formulas (\ref{nbvbvnx}) and (\ref{nvbklkz}).}
\end{table}

\begin{table}[]
\begin{center}
\begin{tabular}{|c|c|c|c|c|} \hline
{\bf Index}& Threshold & Real data & Surrogate data & Confidence level\\ \hline
Dow Jones I. A.& $-18\%$ & $5$ & $0,0,491,387,110,12$ & $\approx 99\%$\\ \hline
S\&P500&$-10\%$ & $7$ & $0,0,183,337,283,130,48,15,4$ &$\approx 99.5\%$\\ \hline
Nasdaq Comp.   & $-14\%$ & $5$ & $478,398,117,7$   & $\approx 99\%$\\ \hline
TSE 300 Comp  .& $-10\%$ & $4$ & $0,642,300,54,4$ & $\approx 99.5\%$ \\ \hline
All Ordinaries & $-14\%$ & $3$ & $0,941,55,3,1$ & $\approx 99.9\%$\\ \hline
Strait Times   & $-29\%$ & $1$ & $1000,0$       & $\approx 100\%$  \\ \hline
Hang Seng      & $-23\%$ & $5$ & $0,0,791,193,13,3$ & $\approx 99.5\%$\\ \hline
Nikkei 225&$-14\%$& $4$ & $0,1,384,354,173,66,18,3,1$& $\approx 75\%$\\ \hline
FTSE 100       & $-23\%$ & $1$ & $996,4$      & $\approx 99.5\%$ \\ \hline
CAC 40         & No & visible  & outliers     & $\approx 0\%$\\ \hline
DAX        & $-13\%$ & $6$ & $0,452,391,122,32,2,1$ & $\approx 99.9\%$\\ \hline
MIBTel         & No & visible & outliers & $\approx 0\% $ \\ \hline \hline
{\bf Currencies}&Threshold&Real data&Surrogate data&Confidence level\\ \hline
DM/US\$    & $-8\%$  & $3$ & $620,305,69,5,1$   & $\approx 99.9\%$  \\ \hline
YEN/US\$   & $-12\%$ & $3$ & $863,137$          & $\approx 100\%$ \\ \hline
CHF/US\$   & $-9\%$  & $3$ & $706,245,45,4$     & $\approx 99.5\%$ \\ \hline
{\bf Gold} & $-17\%$ & $4$ & $799,182,19$        & $\approx 100\%$ \\ \hline
\end{tabular}
\vspace{5mm}
\caption{\label{surrddmarkets} For each of the indices and currencies
named in the first column, a threshold given by the second column
has been chosen by identifying identifying the point of breakdown 
from the stretched exponential,
{\it i.e.}, the crossover point from bulk to outliers in the distribution
(the first point after the crossover point
is then the smallest outlier and the threshold is the integer value
of that drawdown). The
third column gives the number of drawdowns above the threshold in the 
true data. The fourth column gives the number of
surrogate data sets with $0,1, 2, 3, ...$ 
drawdowns larger than the threshold. The last column quantifies
the corresponding confidence level.}
\end{center}
\end{table}

\begin{table}[]
\begin{center}
\begin{tabular}{|c|c|c|c|c} \hline
{\bf Company}& Threshold & True data & Surrogate data&Confidence level\\ \hline
Microsoft (MSFT) &  $-30\%$   & $2$    & $0,777,205,17,1$ & $\approx 78\%$ \\ \hline
Cisco &  $-34\%$   & $1$    & $817,178$ &  $\approx 82\%$  \\ \hline
General Electric (GE)   &  $-29\%$   & $1$    & $969,31$  &  $\approx 97\%$   \\ \hline
Intel (INTC) &  $-33\%$   & $2$    & $739,239,22$ &  $\approx 98\%$  \\ \hline
Exxon-Mobil (XOM)  &  $-18\%$   & $2$    & $1,929,67,3$ &  $\approx 93\%$  \\ \hline
Lucent (LU)   &  $-26\%$   & $2$    & $2,844,140,14$ & $\approx 85\%$  \\ \hline
Wall Mart (WMT)  &  $-22\%$   & $6$    & $236,341,281,107,28,7$ & $\approx 100\%$\\ \hline
IBM  &  $-30\%$   & $1$    & $0,991,9$ &   $\approx 0\%$  \\ \hline
AT\&T (T)    &  $-24\%$   & $4$    & $0,0,0,952,47,1$ & $\approx 95\%$    \\ \hline
Citigroup (C)    &  $-21\%$   & $5$    & $0,1,577,337,69,16$ & $\approx 98.5\%$ \\ \hline
Sun Microsystem (SUNW) &  $-40\%$   & $1$    & $701,251,44,4$ & $\approx 70\%$   \\ \hline
Texas Instruments (TXN)  &  $-36\%$   & $2$    & $0,505,476,19$ & $\approx 50\%$  \\ \hline
Hewlett Packard (HWP)  &  $-29\%$   & $1$    & $506,407,79,8$ & $\approx 50\%$  \\ \hline
SBC Communications (SBC)  &  $-20\%$   & $1$    & $929,68,3$  & $\approx 93\%$ \\ \hline
Merck (MRK)  &  $-20\%$   & $2$    & $775,206,16$ & $\approx 98\%$   \\ \hline
EMC  &  $-39\%$   & $1$    & $0,891,105,4$ & $\approx 0\%$  \\ \hline
Pfizer (PFE)  &$-16\%$ & $6$ & $0,127,290,262,209,77,27,7,1$ & $\approx 96.5\%$\\ \hline
MCI WorldCom (WCOM) & $-22\%$ & $3$ & $31,150,257,251,178,88,31,9,5$ & $\approx 18\%$\\ \hline
Coca Cola (KO)   &  $-29\%$   & $2$    & $0,998,2$  & $\approx 100\%$ \\ \hline
Qualcomm (QCOM) &  $-39\%$   & $1$    & $623,310,61,6$  & $\approx 70\%$ \\ \hline
Appl. Materials (AMAT) &  $-36\%$   & $2$    & $475,390,114,19,2$  & $\approx 87\%$  \\ \hline
Procter\&Gamble (PG)   &  $-37\%$   & $2$    & $0,0,1000$  & $\approx 0\%$  \\ \hline
General Motors (GM)   &  $-30\%$   & $1$    & $776,224$  & $\approx 78\%$  \\ \hline
Medtronic (MDT)  &  $-29\%$   & $1$    & $928,69,3$ & $\approx 93\%$   \\ \hline
Ford (F)    &  $-27\%$   & $1$    & $785,211,4$ & $\approx 78\%$  \\ \hline
\end{tabular}
\vspace{5mm}
\caption{\label{surrddcomp} For each of the companies
named in the first column, a threshold given by the second column
has been chosen by identifying the point of breakdown from the stretched exponential, {\it i.e.}, the crossover point from bulk to outliers in the distribution
(the first point after the crossover point
is then the smallest outlier and the threshold is the integer value
of that drawdown). The
third column gives the number of drawdowns above the threshold in the 
true data. The fourth column gives the number of
surrogate data sets with $0,1, 2, 3, ...$ 
drawdowns larger than the threshold. The last column quantifies
the corresponding confidence level.
}
\end{center}
\end{table}

\end{document}